\documentclass[aps,pra,twocolumn,floatfix, superscriptaddress]{revtex4-2}

\usepackage{graphicx} 
\usepackage{braket}
\usepackage[english]{babel}
\usepackage{amsmath}
\usepackage{placeins}  
\usepackage[ruled,vlined]{algorithm2e}
\usepackage{flushend}

\usepackage{quantikz}
\usepackage[colorlinks=true,
            linkcolor=blue,
            citecolor=blue,
            urlcolor=blue]{hyperref}
\usepackage{subcaption}
\usepackage{physics}
\usepackage{setspace}

\SetKwComment{tcp}{\small\ttfamily$\triangleright$\ }{}

\begin{document}

\title{A Case Study on Noise Resilient Operator Selection in Adaptive Variational Quantum Algorithms}
\author{Soorya Haravu}
\email{soorya.haravu@gmail.com}
\affiliation{Lake Zurich High School, Lake Zurich, IL 60047, USA}

\author{Mafalda Ram\^{o}a}
\email{mafalda@vt.edu}
\affiliation{Department of Physics, Virginia Tech, Blacksburg, VA 24061, USA}
\affiliation{Virginia Tech Center for Quantum Information Science and Engineering,
Blacksburg, VA 24060, USA}

\author{Bharath Sambasivam}
\email{sbharath@vt.edu}
\affiliation{Department of Physics, Virginia Tech, Blacksburg, VA 24061, USA}
\affiliation{Virginia Tech Center for Quantum Information Science and Engineering,
Blacksburg, VA 24060, USA}

\date{\today}

\date{\today}
\begin{abstract}
Hardware noise has been shown to significantly impact the accuracy of ADAPT-VQE, a ground state preparation algorithm. While previous work has studied the impact of noise on its parameter optimization step, its impact on the critical operator selection step remains comparatively unexplored. In this work, we examine the impact of a variety of noise channels on this step, using a linear H$_3$ molecule as a test case. We show that, despite the selection criterion's natural resilience to some noise, both coherent and incoherent noise can prevent convergence for sufficiently high noise rates. We employ quantum error mitigation techniques--dynamical decoupling, zero noise extrapolation, and Pauli twirling--and show that when combined appropriately, these techniques are capable of restoring a successful convergence profile. Our results highlight how error mitigation can improve the performance of ADAPT-VQE and enable convergence in the presence of hardware noise, offering valuable insights into the implementation of the algorithm on near-term quantum hardware.
\end{abstract}
\maketitle

\section{Introduction}

Understanding the quantum-mechanical behavior of molecules offers valuable insight into their chemical properties, which are relevant to many modern industries. In particular, the task of finding the ground state of a molecular system has applications ranging from materials development to modern pharmaceutical production \cite{belaloui_ground-state_2025, cao_potential_2018, golub_eigenvalue_2000}. The ground state of a physical system can be obtained by solving the time-independent Schr\"odinger equation---however, since the computational requirements of such an approach scale exponentially with the size of the system, the method is intractable for moderate to large molecules. Variational methods are a popular class of computational methods for quantum chemistry that use the variational principle to find the lowest-energy state in the space spanned by a parameterized wave function, called \textit{the ansatz}. However, due to the limitations of a classical computer in storing and manipulating the state of a quantum-mechanical system whose full description requires exponential memory, classical ans\"atze have several shortcomings \cite{ramoaAnsatzeNoisyVariational2022,aszaboostlund}.

Quantum computers offer enticing new avenues for tackling this problem by encoding fermionic wave functions in qubit states. In particular, the variational quantum eigensolver (VQE) has emerged as a leading option for this task in the near term, since it leverages the capabilities of quantum computers to use ans\"atze which are classically intractable without requiring long coherent sequences of gates that are currently impractical due to hardware limitations~\cite{peruzzo_variational_2014}. VQE uses a classical computer to minimize a cost function evaluated on a quantum computer. This cost function corresponds to the expectation value of an observable on a state prepared by a parameterized quantum circuit acting on a reference state---often an approximate solution obtained from classical calculations, e.g., the Hartree-Fock ground state for electronic structure problems in chemistry. By taking the Hamiltonian of the system as the observable, the cost function will be the energy---thus, with a suitable ansatz, VQE can in principle find the ground state energy and the corresponding quantum state~\cite{tilly_variational_2022,mcclean_theory_2016, peruzzo_variational_2014}. 

The construction of the ansatz plays a significant role in the effectiveness of VQE; different applications require different ans\"atze. Even within the same type of problem, we may find that the precision achievable by a certain ansatz varies from instance to instance. ADAPT-VQE is a well-known algorithm that tailors the ansatz specifically to the system at hand, leading to a more efficient state preparation and increased noise resilience compared to static VQEs~\cite{grimsley_adaptive_2019, ramoa_reducing_2025,tang2021qubitadaptvqe,yordanov_qubit-excitation-based_2021, dalton_quantifying_2024}. This algorithm constructs the ansatz dynamically, selecting operators from a pool based on local gradient measurements that are performed on the quantum computer ~\cite{anastasiou_how_2023}. ADAPT-VQE has seen orders-of-magnitude improvements in relevant metrics (such as gate counts and measurement costs) since it was proposed, with remarkable progress in the hardware efficiency of the operator pool, as well as the variational circuit~\cite{ramoa_reducing_2025,yordanov_qubit-excitation-based_2021,tang2021qubitadaptvqe,anastasiouTETRISADAPTVQEAdaptiveAlgorithm2022,ramoa2026codesignedadaptivequantumstate}, measurement cost requirements~\cite{stadelmann2025strategiesovercominggradienttroughs,ramoaReducingMeasurementCosts2024b,anastasiou_how_2023}, and generalization of the algorithm to applications such as thermal states, lattice models, periodically driven systems, or combinatorial optimization \cite{sambasivamTEPIDADAPTAdaptiveVariational2025,dykeScalingAdaptiveQuantum2023,kumarFloquetADAPTVQEQuantumAlgorithm2025,zhuAdaptiveQuantumApproximate2022}. 

Despite all the algorithmic improvements, the levels of hardware noise present in today's quantum computers remain an impediment to the experimental implementation of ADAPT-VQE. Recent studies have shown that under some assumptions, incoherent noise models prevent ADAPT-VQE from reaching chemical accuracy (the typical target precision for chemical systems relevant for wet-lab experiments)~\cite{mukherjee_comparative_2023, long_layering_2024}. While hardware noise is often seen as a hurdle, research has shown that stochastic hardware noise can be beneficial to variational quantum algorithms such as ADAPT-VQE by preventing the optimization process from getting stuck on strict saddle points \cite{liu2024stochastic}. Despite these potential advantages, hardware noise remains an obstacle to reaching chemical accuracy. Various methods have been developed specifically to improve the noise resilience of ADAPT-VQE. One such method, known as TETRIS or layering, adds multiple operators to the ansatz in each iteration, in an effort to produce a more compact circuit with fewer idle qubits at any given time \cite{anastasiouTETRISADAPTVQEAdaptiveAlgorithm2022}. This has been shown to improve the resilience of ADAPT-VQE when affected by amplitude damping or phase damping, but not when affected by depolarizing noise~\cite{long_layering_2024}.

Error mitigation (EM) methods are more generic approaches that mostly employ post-processing techniques to reduce the impact of errors on the final expectation values. An example is zero noise extrapolation (ZNE), which has been shown to effectively mitigate gate-level amplitude damping, phase damping, and depolarizing noise within both ADAPT-VQE frameworks and the wider VQE field~\cite{mukherjee_comparative_2023, pelofske_increasing_2024}. Dynamical decoupling (DD), another technique that performs pulses that effectively implement the identity operation on idle qubits, can be combined with ZNE to further improve the results~\cite{weaving_contextual_2025, dalton_quantifying_2024, mukherjee_comparative_2023}. 

Many studies have researched the effect of noise on ADAPT-VQE~\cite{mukherjee_comparative_2023, long_layering_2024, dalton_quantifying_2024, anastasiouTETRISADAPTVQEAdaptiveAlgorithm2022}. However, these studies exclusively focus on the impact of incoherent noise. Coherent noise can often be more damaging to quantum algorithms as it scales infidelity quadratically rather than linearly like incoherent noise~\cite{IversonPreskill2020CoherenceLogicalChannels}. Therefore, coherent noise has the potential to represent a significant impediment to the effectiveness of ADAPT-VQE, motivating the study of its impact on the algorithm. Additionally, previous work focuses exclusively on the impact of noise on the energy evaluation and optimization process of ADAPT-VQE, neglecting operator selection. The efficacy of the algorithm depends on the precision of the operator selection process, since this guides the ansatz construction. Noise-induced errors during selection may misdirect the trajectory of the algorithm, resulting in suboptimal anz\"atze and inaccurate energy estimates. Understanding these effects is therefore essential to predict the algorithm's performance on near-term hardware. This study aims to address these critical gaps in the literature by diving into the impact of noise on operator selection and studying which EM can mitigate it and to what degree. Our findings demonstrate that both coherent and incoherent noise significantly distort gradient distributions, obscuring the signal needed to identify the optimal operator. This distortion often leads to the construction of a suboptimal ansatz, ultimately resulting in an overestimation of the ground state energy. We show that we are able to counter these effects using EM strategies, and employ numerical simulations to understand which methods are most effective. We observe that the combination of DD with ZNE is consistently the most robust against incoherent noise, while coherent noise is best managed by pairing PT with ZNE, DD, or a combination of both (PT+DD+ZNE).

This paper is structured as follows. Sec.~\ref{Section II}, the background section, focuses on the details of the VQE and ADAPT-VQE algorithms. Sec.~\ref{Section III} discusses the methods we employed in this work, including details about the noise channels we consider and the EM techniques we implement. Sec.~\ref{Section IV} contains numerical simulation results. Sec.~\ref{Section IV A} discusses the impact of incoherent noise on ADAPT-VQE---particularly, its impact on gradient magnitude and operator selection. Sec.~\ref{Section IV B} presents a similar discussion for coherent noise. Finally, Sec.~\ref{Section V} contains concluding remarks.


\section{Background}\label{Section II}

The VQE algorithm is designed to prepare eigenstates of physical systems whose evolution is governed by a Hamiltonian. This involves solving the time-independent Schr\"odinger equation,
\begin{equation}\label{equation1}
    \hat{H}\ket{\psi} = E \ket{\psi},
\end{equation}
where $\hat{H}$ is the problem Hamiltonian~\cite{aszaboostlund}. The solutions to Eq.~\eqref{equation1} are the eigenstates of $\hat{H}$, $\ket{\psi_i}$, with corresponding eigenvalues $E_i$. We will focus on finding the ground state $\ket{\psi_0}$ and its energy $E_0$. The fundamental basis for VQE is the variational principle of quantum mechanics, which states that 

\begin{equation}\label{equation2}
    E_0 \leq \frac{\bra{\psi}\hat H\ket{\psi}}{\braket{\psi|\psi}}.
\end{equation}

The right side of Eq.~\eqref{equation2} is the Rayleigh-Ritz quotient; the variational principle can approximate $E_0$ by minimizing this quotient. VQEs define a parameterized wave function $\ket{\psi(\vec{\theta})}$ and minimize the energy by adjusting the parameter vector $\vec{\theta}$ in an effort to find the true ground state of $\hat{H}$~\cite{ramoa_reducing_2025, anastasiou_how_2023, romero_strategies_2018, tilly_variational_2022, peruzzo_variational_2014}. The parameterized wave function $\ket{\psi(\vec{\theta})}$, known as the ansatz, determines the search space in which the algorithm will look for the ground state. In classical computers, the number of bits required to store a generic electronic wave function grows exponentially with the number of orbitals considered, making classical methods impractical for complex molecular systems---but if we use a quantum computer to represent the state, the number of qubits required scales linearly, due to quantum entanglement. Therefore, by preparing the ansatz using a quantum computer and optimizing the variational parameters with a classical computer, VQE has the potential to significantly reduce the computational cost of finding the ground state as compared to classical methods. Additionally, the use of classical optimization makes VQE a hybrid quantum-classical protocol. In contrast with fully quantum algorithms, this avoids deep gate sequences, making the algorithm more suitable for near-term hardware~\cite{peruzzo_variational_2014}. Various VQE ans\"atze have been proposed for molecular electronic structure problems in chemistry. Some of the most well-known examples include unitary coupled cluster singles and doubles (UCCSD) ans\"atze~\cite{grimsleyTrotterizedUCCSDAnsatz2020, Rattew2019EVQE} and hardware-efficient ans\"atze (HEA)~ \cite{kandala_hardware-efficient_2017, Wang2023EntanglementVariationalHEA}; however, both face significant challenges that limit their efficiency.

While convenient to implement, HEAs are prone to barren plateaus, a problem where the gradient landscape of the cost function becomes so featureless that optimization effectively stalls unless exponentially many measurements are used to resolve the search direction~\cite{McClean2018BarrenPlateaus}. This issue stems from the fact that these ans\"atze seek to explore the entire Hilbert space without any  problem-specific guidance. The lack of a targeted search area leads the gradient of the cost function with respect to the variational parameters to vanish exponentially with the size of the system, rendering the optimizer unable to navigate the flat energy landscape for moderate to large problems~\cite{McClean2018BarrenPlateaus}. This limitation highlights the need for a more targeted, adaptive construction like the one provided by ADAPT-VQE, which avoids barren plateaus~\cite{Grimsley2023barrenplateausADAPT}.

UCCSD ans\"atze also suffer from various problems that stem from a system-agnostic construction. This method uses all single and double excitation operators when only a system-specific subset is truly necessary to reach the ground state~\cite{grimsley_adaptive_2019}. Furthermore, implementing the ansatz with a quantum circuit requires a choice of operator ordering that leads to energy variations that are relevant on a chemical scale, i.e., the ansatz is not chemically well-defined~\cite{grimsley_is_2020}. Additionally, UCCSD ans\"atze are known to fail to reach chemical accuracy for strongly correlated molecular systems, even if the system is small~\cite{grimsley_adaptive_2019}.

These limitations of UCCSD and HEA are precisely the reason why ADAPT-VQE was developed~\cite{grimsley_is_2020}. In contrast to the fixed structure of the UCCSD ansatz, ADAPT-VQE employs an iterative construction process. This approach produces a problem-specific ansatz that eliminates redundant operators, ensuring that the ansatz is more noise-resilient and parameter-efficient~\cite{grimsley_is_2020}. A key part of ADAPT-VQE is the operator pool, a set of anti-Hermitian operators from which ADAPT-VQE will choose generators to construct the ansatz. The operator pool allows the algorithm to narrow down the search space for potential solutions, avoiding the effects of barren plateaus~\cite{Grimsley2023barrenplateausADAPT}. The pseudo-code for ADAPT-VQE~\cite{ramoaReducingMeasurementCosts2024b} can be found in Alg.~\ref{alg:ADAPT-VQE}, where $\epsilon$ and $L$ are hyperparameters that define the termination of ADAPT-VQE. The algorithm terminates if the gradient norm is below $\epsilon$, or if the maximum number of iterations is greater than $L$, whichever happens first.

\begin{algorithm*}[t]
\setstretch{1.4}
\SetKwInOut{Input}{Input}
\SetKwInOut{Output}{Output}

\newcommand{\commfont}[1]{\textnormal{#1}}

\Input{$\ket{\psi_{\text{ref}}}, \{\hat{A}_i\}_{i=1}^N, \hat{H}$ \tcp*[r]{\commfont{Problem specification}}}
$\epsilon, L$ \tcp*[r]{\commfont{Hyperparameters}}
\Output{$\ket{\psi^*}, \vec{\theta}^*, E^*$}

$k \leftarrow 0$\;
$\vec{\theta}_0 \leftarrow \{\}$\;
$E_{\text{ref}} \leftarrow \text{measure energy}\!\left(\hat{H}, \ket{\psi_{\text{ref}}}\right)$\;

\While{$k < L$}{
    $k \leftarrow k + 1$\;

    \For{$i \leftarrow 1 \dots N$}{
        Measure $g_i = \left. \frac{\partial E}{\partial \theta_i} \right|_{\theta_i=0} = \bra{\psi_{k-1}} \left[ \hat H, \hat A_i \right] \ket{\psi_{k-1}}$ \tcp*[r]{\commfont{Measure gradients}}
    }

    $i^* \leftarrow \arg\max_{i} |g_i|$ \tcp*[r]{\commfont{Select new operator}}

    \vspace{0.5em}
    $G \leftarrow \| \{g_1, \dots, g_N\} \|_F$ \tcp*[r]{\commfont{Calculate gradient norm}}

    \If{$G > \epsilon$}{
        $\ket{\psi_k} \leftarrow e^{\theta_{i^*}\hat A_{i^*}} \ket{\psi_{k-1}}$ \tcp*[r]{\commfont{Grow ansatz}}

        $\vec{\theta}_k \leftarrow \{\vec{\theta}_{k-1}, 0\}$ \tcp*[r]{\commfont{Grow parameter vector}}

        $E_k, \vec{\theta}_k \leftarrow \text{VQE}(\hat{H}, \ket{\psi_k}, \vec{\theta}_k)$ \tcp*[r]{\commfont{Minimize energy}}
    }
    \Else{
        \Return $\ket{\psi_{k-1}}, \vec{\theta}_{k-1}, E_{k-1}$ \tcp*[r]{\commfont{Successful convergence}}
    }
}

\Return $\ket{\psi_k}, \vec{\theta}_k, E_k$ \tcp*[r]{\commfont{Convergence target unmet}}

\caption{ADAPT-VQE}
\label{alg:ADAPT-VQE}
\end{algorithm*}

The choice of operator pool $ \{\hat{A}_i\}_N$ is important to the performance of ADAPT-VQE. Different pools define different variants of ADAPT-VQE, with the most hardware-efficient option for fermionic systems, including the electronic structure problem, currently being CEO-ADAPT-VQE~\cite{ramoa_reducing_2025}, followed by QEB-ADAPT-VQE \cite{yordanov_qubit-excitation-based_2021}, and Qubit-ADAPT-VQE \cite{tang2021qubitadaptvqe}. 

Typically, the Hartree-Fock state is used as the initial reference state for electronic structure problems $(\ket{\psi_{\text{ref}}} = \ket{HF})$. After each iteration, a chosen operator is appended to the ansatz with its variational parameter initialized to 0~\cite{ramoaReducingMeasurementCosts2024b}. 

It is important to note that adding operators to the ansatz does not drain the pool, as each operator can be selected multiple times, albeit not consecutively (since immediately after an operator is selected, the corresponding energy derivative will be zero). At the $k^{\text{th}}$ iteration, the ansatz will have the form

\begin{equation}\label{equation8}
\ket{\psi_k} = e^{\theta_k A_k} \cdots e^{\theta_1 A_1} \, \ket{\psi_{\text{ref}}}.
\end{equation}

Operator selection works by choosing the pool operator with the highest potential to minimize the cost function, i.e., the pool operator with the largest gradient magnitude. Using the fact that pool operators are anti-hermitian, the gradient associated with any pool operator $A_i$ can be calculated using the equation \cite{grimsley_adaptive_2019}

\begin{align}
\left. \frac{\partial E}{\partial \theta_i} \right|_{\theta_i = 0}
&= \left. \frac{\partial}{\partial \theta_i}
\bra{\psi_{k-1}}
e^{-\theta_i \hat A_i} \hat H e^{\theta_i \hat A_i}
\ket{\psi_{k-1}} \right|_{\theta_i = 0} \notag \\
&= \bra{\psi_{k-1}} [\hat H, \hat A_i] \ket{\psi_{k-1}}.
\label{eq: equation9}
\end{align}

It should be noted that the operator subscripts in Eq.~\eqref{equation8} denote the order in which they were appended to the ansatz, whereas in Eq.~\eqref{eq: equation9} the subscript $i$ denotes any pool operator that we consider to append to the ansatz.

We can formulate Eqs.~\eqref{equation8} and \eqref{eq: equation9} more generically using density matrices, which allows us to use these formulas for states suffering from incoherent noise:

\noindent Defining
\[
\rho_{k} := \hat U_{k}\,\rho_{0}\,\hat U_{k}^{\dagger},\qquad
\hat U_{k} := e^{\theta_{k}\hat A_{k}}\cdots e^{\theta_{1}\hat A_{1}},
\]
we have

\begin{align}\label{equation11}
\begin{split}
\left.\frac{\partial E}{\partial \theta_i}\right|_{\theta_i=0}
&= \left.\frac{\partial}{\partial \theta_i}\,
\mathrm{Tr}\!\left[\hat H\, e^{\theta_i \hat A_i}\,\rho_{k}\, e^{-\theta_i \hat A_i}\right]\right|_{\theta_i=0}\\
&= \mathrm{Tr}\!\left([\hat H,\hat A_i]\,\rho_{k}\right).
\end{split}
\end{align}
Using Eq.(\ref{equation11}), the operator selection process can be modified to include the impact of noise, as shown in Alg.~\ref{alg:noisy-gp-select}, where $i^*$ is the chosen operator index to be appended to the ansatz and $\mathcal{E}$ represents the selected noise channel. $\mathcal{E}$ is applied to the state before the gradient is evaluated for a given operator, allowing it to distort gradient values and potentially alter which operator is appended to the ansatz at a given iteration. 
\begin{algorithm*}[t]
\setstretch{1.4}
\SetKwInOut{Input}{Input}
\SetKwInOut{Output}{Output}

\newcommand{\commfont}[1]{\textnormal{#1}}

\Input{
$\rho_{k-1}, \{\hat{A}_i\}_{i=1}^N, \hat{H}, \mathcal{E}$ \tcp*[r]{\commfont{Problem specification}}
$\epsilon$ \tcp*[r]{\commfont{Hyperparameter}}
}

\Output{$i^*$}

\For{$i \leftarrow 1 \dots N$}{
    Measure $g_i \leftarrow  
    \left. \dfrac{\partial E}{\partial \theta_i} \right|_{\theta_i=0}
    = \mathrm{Tr}\!\left([\hat H,\hat A_i]\,\mathcal{E}(\rho_{k-1})\right)$ \tcp*[r]{\commfont{Measure noisy gradients}}
}

$i^* \leftarrow \arg\max_{i} |g_i|$ \tcp*[r]{\commfont{Return operator of selected index}}

\Return{$i^*$}

\caption{Noisy operator selection for ADAPT-VQE}
\label{alg:noisy-gp-select}
\end{algorithm*}

\section{Methods}\label{Section III}

\subsection{Simulation details and test molecules} \label{Section III A}
We perform classical simulations to evaluate the performance of ADAPT-VQE under different noise models and error mitigation strategies. A molecule with three hydrogen atoms in linear geometry (henceforth denoted H$_3$) with interatomic distance 1.5\AA{} was chosen as the test system. Using the minimal STO-3G basis set, this molecule can be represented by 6 qubits. As the costs of  noisy simulation and error mitigation can grow quickly with the number of qubits, this system size achieves a convenient balance between simulation time and a nontrivial run of ADAPT-VQE. To eliminate statistical fluctuations, all simulations are performed with no sampling noise, as has been the case in most ADAPT-VQE simulations in the literature \cite{grimsley_adaptive_2019,yordanov_qubit-excitation-based_2021,tang2021qubitadaptvqe,ramoa_reducing_2025}.

While the state-of-the-art version of the algorithm is CEO-ADAPT-VQE \cite{ramoa_reducing_2025}, we choose to simulate QE-ADAPT-VQE due to the fact that CEO-ADAPT-VQE adds up to three operators to the ansatz per iteration, which complicates the analysis of the effect of hardware noise on the operator selection. We note that CEOs consist of linear combinations of QEs, which in turn consist of fermionic excitations under the Jordan-Wigner transform after the removal of the anticommutation string (consisting of Pauli-$Z$ operators). Hence, we expect the results in this paper to readily generalize to other choices of operator pools.

\subsection{Noise Models}\label{Section III B}

The impact of hardware noise on parameter optimization has been studied in the context of VQE algorithms, as parameter optimization for expectation value extremization is not specific to ADAPT-VQE \cite{tilly_variational_2022, kandala_hardware-efficient_2017}. Therefore, the focus of this paper is to study the impact of noise on the operator selection step of ADAPT-VQE by injecting noise \emph{exclusively} into the gradient measurements used for the operator selection process. 

To test ADAPT-VQE under various noisy conditions, we simulate seven distinct single-qubit noise channels. Four of these models are incoherent, including amplitude damping, phase damping, and depolarizing noise channels, as well as a combination of all three, referred to as `all incoherent'. By studying the individual contributions of each incoherent channel, we can understand which types of noise most significantly impact ADAPT-VQE. However, because real noise channels combine multiple types of noise, the simultaneous action of all three incoherent channels can give a more realistic insight into the effect of incoherent noise on the operator selection in ADAPT-VQE. 

The amplitude damping, phase damping, and depolarizing noise channels can be characterized by their Kraus representation,
\begin{equation}\label{equation12}
\mathcal{E}(\rho) = \sum_{k} K_{k}\,\rho\,K_{k}^{\dagger}, 
\end{equation}
where $\rho$ is the density matrix of the initial quantum state and $k$ indexes the Kraus operators of the channel $\mathcal{E}$, which satisfy the trace-preserving condition ($\sum_{k} K_{k}^{\dagger} K_{k} = I$, where $I$ is the identity matrix). In order to simulate noise, we convert the fermionic ansatz to a circuit of fundamental basis gates via the Jordan-Wigner mapping. From there, noise can be implemented after each unitary basis gate using the following equation

\begin{equation}\label{equation13}
\rho_f = \sum_{k} K_{k}\, U \,\rho\, U^{\dagger} K_{k}^{\dagger},
\end{equation}
where $U$ is a unitary gate in the ansatz circuit and $\rho_f$ is the noisy quantum state (after the action of the noisy unitary)

Amplitude damping, phase damping, and depolarizing noise are defined by different Kraus operators that act on a quantum state. The Kraus operators for amplitude damping (AD) are

\begin{equation}
K^{AD}_0 =
\begin{pmatrix}
1 & 0\\[2pt]
0 & \sqrt{1-\gamma}
\end{pmatrix},
\qquad
K^{AD}_1 =
\begin{pmatrix}
0 & \sqrt{\gamma}\\[2pt]
0 & 0
\end{pmatrix},
\end{equation}
with the damping probability $\gamma \in [0,1]$. 

Using the same notation, the Kraus operators for phase damping (PD) can be defined as

\[
K^{PD}_0  =
\begin{pmatrix}
1 & 0\\
0 & \sqrt{1-\gamma}
\end{pmatrix},
\qquad
K^{PD}_1 =
\begin{pmatrix}
0 & 0\\
0 & \sqrt{\gamma}
\end{pmatrix},
\]
where $\gamma$ determines the exponential decay of coherence and is also $\in [0,1]$.

Unlike amplitude and phase damping, depolarizing noise (DN) has four Kraus operators. We can represent them in terms of Pauli operators as

\begin{equation}
\begin{aligned}
K^{\mathrm{DN}}_{0}&=\sqrt{1-\tfrac{3\gamma}{4}}\,I,  &\qquad
K^{\mathrm{DN}}_{1}&=\sqrt{\tfrac{\gamma}{4}}\,X,\\[4pt]
K^{\mathrm{DN}}_{2}&=\sqrt{\tfrac{\gamma}{4}}\,Y,  &\qquad
K^{\mathrm{DN}}_{3}&=\sqrt{\tfrac{\gamma}{4}}\,Z,
\end{aligned}
\end{equation}
where $I, X, Y, Z$ are the single-qubit Pauli operators, and $\gamma$ is the depolarizing probability. Note that, for consistency, we always use $\gamma$ to label parameters related to noise strength in incoherent noise channels.

In addition to the incoherent noise channels, we consider three coherent ones, consisting of over-rotations around the X or Z axis at each gate, or of a combination of both that we refer to as `all coherent'. 

In order to simulate the effects of over-rotation errors on the ansatz, we apply an additional $R_P$ gate to all single-qubit gates in the ansatz circuit, defined as
\[ 
R_P = e^{-i\frac{\phi}{2} P},\]
where $\phi$ is the rotation angle and $P$ is a Pauli gate (in our case, $P\in\{X,Z\}$). 

Testing the effect of coherent noise channels like the $X$ and $Z$ over-rotation errors on ADAPT-VQE is very important, as coherent noise is able to cause the infidelity of a quantum circuit to increase quadratically with circuit size, rather than linearly (at worst) like incoherent noise \cite{IversonPreskill2020CoherenceLogicalChannels}. 

For a comprehensive analysis, we consider several values for the strength of the noise channels $\gamma$. In the realm of incoherent noise channels, $\gamma$ values are usually set lower than 0.01 to model realistic noise on NISQ devices \cite{tang2021qubitadaptvqe, Bilkis2024VAns, Javanmard2022DynaPhaseTransitions}. However, far higher values of  $\gamma$ have been tested ($0.3 < \gamma < 0.5$) \cite{Martinez2019Entropy}. For coherent noise, $\phi$ values up to $0.1 \pi$ are generally considered appropriate \cite{Bravyi2018CorrectingCoherentErrors}.  Based on this, we consider 3 noise levels for both $\gamma $ and $\phi$ - 0.01, 0.1, and 0.3 - for all coherent and incoherent noise channels. The purpose is to analyze three distinct regimes: one where ADAPT-VQE succeeds without EM; one where ADAPT-VQE does not succeed without EM, but might succeed with it; and one where the noise is so strong that ADAPT-VQE is very unlikely to succeed even with EM. We test the algorithm under all these regimes to offer a well-rounded and complete view of the efficacy of EM in the selection of operators in ADAPT-VQE.

\subsection{Quantum Error Mitigation Methods }\label{Section III C}

Three types of error mitigation methods with low or no resource overhead were used to mitigate the effects of noise on ADAPT-VQE operator selection: dynamical decoupling (DD),  zero noise extrapolation (ZNE), and Pauli twirling (PT). 

DD is an error suppression strategy designed to prevent decoherence in spin systems. The method works by applying controlled electromagnetic pulses to the system; in this case,  an $XY$ pulse was implemented into the ansatz circuit due to its effectiveness in a greater variety of systems compared to similar pulse sequences like $XX$ \cite{Tyryshkin2010DDErrors, Souza2011RobustDD}. Pulses were scheduled uniformly across the idle space for each qubit simply for ease of simulation. For more information on scheduled pulse sequences, we refer to Ref.~\cite{ibm_scheduling_passes}.

ZNE is an error mitigation strategy designed to predict a noise-free expectation value of a given observable by scaling up processor noise and extrapolating the noise-free expectation value via regression. This work uses ZNE to extrapolate the noiseless expectation value of the pool gradients.

Noise is often scaled up via a process known as unitary folding, in which a gate $U$ is replaced as 
\begin{equation}
    U \;\to\; U (U^{\dagger} U)^n,
\end{equation}
where  $n$ is a positive integer. By the unitarity of $U$, we have $(U^{\dagger} U)^n=I$; as such, in an ideal simulation, this transformation would not affect the function of the gate. However, since the number of physical operations scales by $\lambda = 1 + 2n$, the result will be more severely affected by noise for larger $n$ if we execute this on a noisy quantum device. Hence, the expectation value of a given observable, in this case the pool gradient $g$, will vary as $\lambda$ increases \cite{giurgica-tiron2020digital, ramoa_reducing_2025, yordanov_qubit-excitation-based_2021,tang2021qubitadaptvqe}. CNOT gates are typically associated with significantly higher errors than single-qubit gates \cite{dalton_quantifying_2024}, and recent studies have proven that folding only CNOT gates can scale noise both efficiently and adequately enough for ZNE \cite{he2020resource}.  Therefore, our study will rely exclusively on folding CNOTs in the ansatz circuit. We can model $g$ as a function of $\lambda$ by measuring it at $m$ different values of $\lambda$.
From there, the noiseless expectation value or $g(0)$ can be extrapolated \cite{kurita_synergetic_2023}. While unitary folding naturally produces odd integer scale factors ($\lambda$ = 1, 3, 5, \dots), this study selects $\lambda \in \{1,2,3,4,5,6\}$. To implement an even scale factor $\lambda$, we employ a stochastic mixture of folding levels \cite{giurgica-tiron2020digital}. Individual CNOTs are randomly assigned a folding order of either $n_{low} = \frac{\lambda-2}{2}$ or $n_{high} =n_{low} +1 = \frac{\lambda}{2}$.

This ensures that every individual CNOT is folded by a different random factor, but the noise level for the circuit as a whole is approximately scaled by the desired even value of $\lambda$. To mitigate the variance introduced by the probabilistic nature of local folding at even scale factors, the reported expectation values for $\lambda \in \{2,4,6\}$ represent the mean of $s$ randomized circuit executions. To remain consistent with the literature, an $s=10$ was used \cite{pelofske_increasing_2024}. 

The method of extrapolation is an important factor that may affect the effectiveness of ZNE.  We employed polynomial extrapolation as it predicts noise-free expectation values more precisely than methods such as linear and Richardson extrapolation, while being less computationally expensive than exponential extrapolation (compared to which it is only marginally less precise) \cite{giurgica-tiron2020digital}. For more information on alternative extrapolation methods, we refer to Ref.~\cite{giurgica-tiron2020digital}. Polynomial extrapolation works by representing the relationship between the noise scaling parameter $\lambda$ and the chosen expectation value $g_d(\lambda)$ as
\begin{equation}
    g_d(\lambda) = c_{0} + c_{1}\lambda + \cdots + c_{d}\lambda^{d}.
\end{equation}

In this case,  $d$ is the order of the polynomial, and $c_0, c_1 ... c_d$  are a set of $d+1$ unknown real parameters. The noise-free expectation value can be approximated by solving for $g_d(0)$ \cite{giurgica-tiron2020digital}. We chose $d=2$ in order to provide an accurate fit for the available data while also preventing overfitting.

For incoherent channels, DD and ZNE have been proven to be effective, both independently and in combination, in mitigating the effects of noise, including in the context of VQE algorithms~\cite{weaving_contextual_2025, giurgica-tiron2020digital, mukherjee_comparative_2023, dalton_quantifying_2024}. In light of this, we consider the parallel application of (1) DD, (2) ZNE, and (3) DD \& ZNE to the noisy ADAPT-VQE simulations, with the aim of comparing the effectiveness of the three approaches in mitigating incoherent noise.

Coherent noise channels are more challenging to deal with than incoherent channels: They have a more severe worst-case scaling of the infidelity with circuit size (as discussed in Sec.~\ref{Section III B} or the Noise Models section), and they are also more difficult to characterize. Turning coherent channels into incoherent (also called stochastic) channels is a significant area of research \cite{winick_concepts_2022}. Pauli twirling (PT), also known as randomized coupling, is a proven method for converting a coherent noise channel into a stochastic channel \cite{winick_concepts_2022, kurita_synergetic_2023, dalton_quantifying_2024}. In our case, PT works by applying random Pauli operators $(X,Y,Z,I)$ around each CNOT gate as

\begin{center}
\begin{quantikz}
\lstick{} & \gate{P_i} & \ctrl{1} & \gate{P_j} & \qw \\
\lstick{} & \gate{P_k} & \targ{}  & \gate{P_l} & \qw ,
\end{quantikz}
\end{center}

\noindent where ${P_i, P_j, P_k, P_l}$ are randomly selected Pauli operators that preserve the logical CNOT operation in each individual circuit. The process of random Pauli selection is repeated $k$ times, and the expectation value is taken as the average over the $k$ circuits. This transforms the coherent channel into an incoherent one \cite{winick_concepts_2022}. Similar to Ref.~\cite{winick_concepts_2022}, we take $k=10$. For real-world NISQ devices, measurement overhead is a significant concern. PT is advantageous over other techniques because it distributes the shot budget across the $k$ randomly selected circuits, meaning there is no meaningful increase in measurement overhead~\cite{wallman_noise_2016}. 

While PT removes the structure of the noise, improving the worst-case scaling, it does not address noise beyond that. Hence, it should be paired with EM methods aimed at mitigating incoherent noise. ZNE and DD are proven choices to synergize with PT \cite{kurita_synergetic_2023, dalton_quantifying_2024}. As there is no consensus on whether PT should be paired with DD or with ZNE, we consider both options independently as well as together \cite{kurita_synergetic_2023, dalton_quantifying_2024}. Table~\ref{tab:placeholder} shows a complete list of all noise models used for coherent and incoherent noise.

\begin{table}[h]
    \centering

    \begin{tabular}{|c|c|}\hline
         Incoherent Noise& Coherent Noise\\\hline
         DD& PT \& ZNE\\\hline
         ZNE& PT \& DD\\\hline
         DD \& ZNE& DD \& PT \& ZNE\\ \hline
    \end{tabular}
\caption{List of EM methods employed to mitigate coherent and incoherent noise models.}
\label{tab:placeholder}

\end{table}
\section{Results}\label{Section IV}

As discussed in the previous section, we aim to analyze the impact of noise and error mitigation techniques on the operator selection step of ADAPT-VQE. This requires a careful analysis of what is meant by `impact'. In what concerns energy evaluations, a typical target is chemical accuracy---this is typically taken as the target regime, including when assessing performance under noise \cite{dalton_quantifying_2024}. However, the same criterion does not apply to gradient evaluations. Naively, if we aim to select the operator with the highest gradient, we need to evaluate the gradients with sufficient accuracy to resolve the difference between the highest and second-highest gradients. There are multiple problems with this perspective. First, we may have degeneracies, i.e., multiple operators with the same gradient, such that this difference vanishes. Second, we do not need to select the operator with the highest gradient to observe good convergence; in general, operators with a high (not necessarily the highest) gradient will significantly impact the energy. As the gradient is a heuristic selection criterion, it is unnecessary to spend resources ensuring that the selected operator is the one with the highest gradient. Third, robust operator selection under noise depends less on maintaining the numerical precision of gradient magnitudes and more on maintaining the overall ordering of gradient magnitudes. Hence, analyzing the impact of noise on operator selection requires careful handling of gradient degeneracies, gradient differences, and the leeway inherent to the heuristic nature of the selection criterion.

Our approach to the analysis of these factors is two-fold: We consider heatmaps with the gradient magnitudes along the evolution of ADAPT-VQE with and without noise or error mitigation, marking the highest one in each iteration; and in parallel, we consider plots of the evolution of the ADAPT-VQE energy error. With the first type of plots, we can observe the gradients of the selected operators and compare them with the noiseless case; with the second type, we can observe the impact of selecting different operators on the performance of the algorithm. With this approach, we can observe the impact of noise and error mitigation techniques on each particular operator selection, as well as on the entire algorithm from a broader perspective. 

Plot styles are standardized for all heatmaps and energy error plots in both the incoherent (Sec.~\ref{Section IV A}) and coherent noise (Sec.~\ref{Section IV B}) sections. Again, we use heatmaps to visualize the gradient landscape (i.e., the distribution of operator gradients at each iteration). A red star marks the operator added to the ansatz in each iteration (i.e., the one with the highest measured gradient). Heatmaps require that every operator in the pool be given a label. For clarity, each of the 24 unique operators in the used QE pool will be consistently assigned an index between 0 and 23. To optimize visual clarity and focus on the regions of greatest dynamical change, the iteration count for heatmaps in this study was tailored to the convergence behavior observed in our specific test cases ($N=10$ for non-EM runs; $N=20$ for EM runs).
Energy error plots show the absolute difference between the recorded energy at each iteration and the true ground-state energy of H$_3$. Iteration 0 is defined as the Hartree-Fock energy, or the baseline energy before the initial ADAPT-VQE run. The chemical-accuracy region---defined by errors below $1.596 \times 10^{-3}$ Hartree, a practical target enabling reliable predictions---is shaded in blue. Like the heatmaps, energy error plots for runs without EM will show only 10 iterations, while those with EM will show 20.

\subsection{Incoherent Noise}\label{Section IV A}

\begin{figure*}[t]
  \centering
  \begin{subfigure}[t]{0.49\textwidth}
    \centering
    \includegraphics[width=\linewidth]{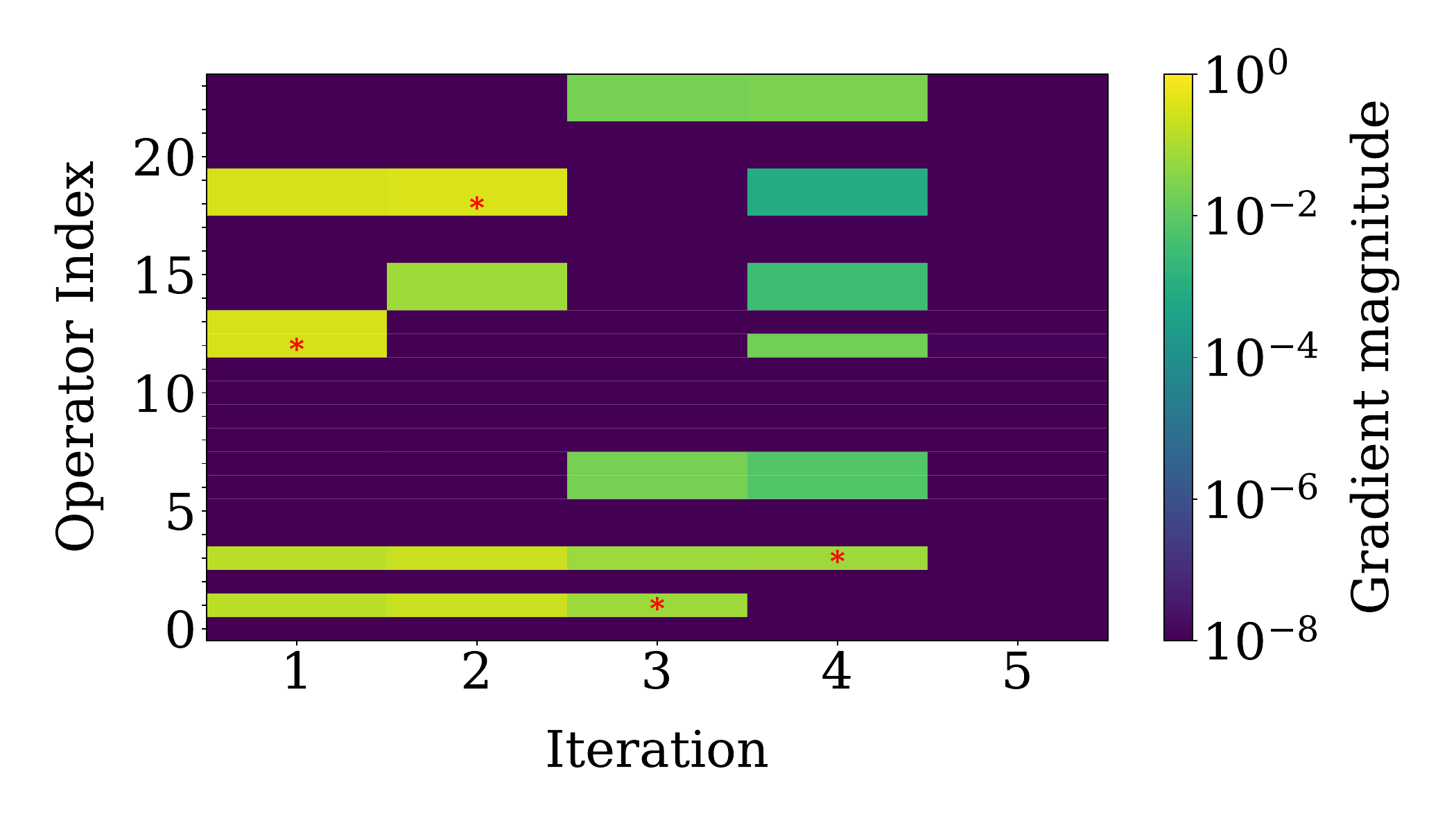}
    \caption{\centering Gradient heatmap}
    \label{fig:Ideal_Heatmap}
  \end{subfigure}%
  \hfill%
  \begin{subfigure}[t]{0.49\textwidth}
    \centering
    \includegraphics[width=\linewidth]{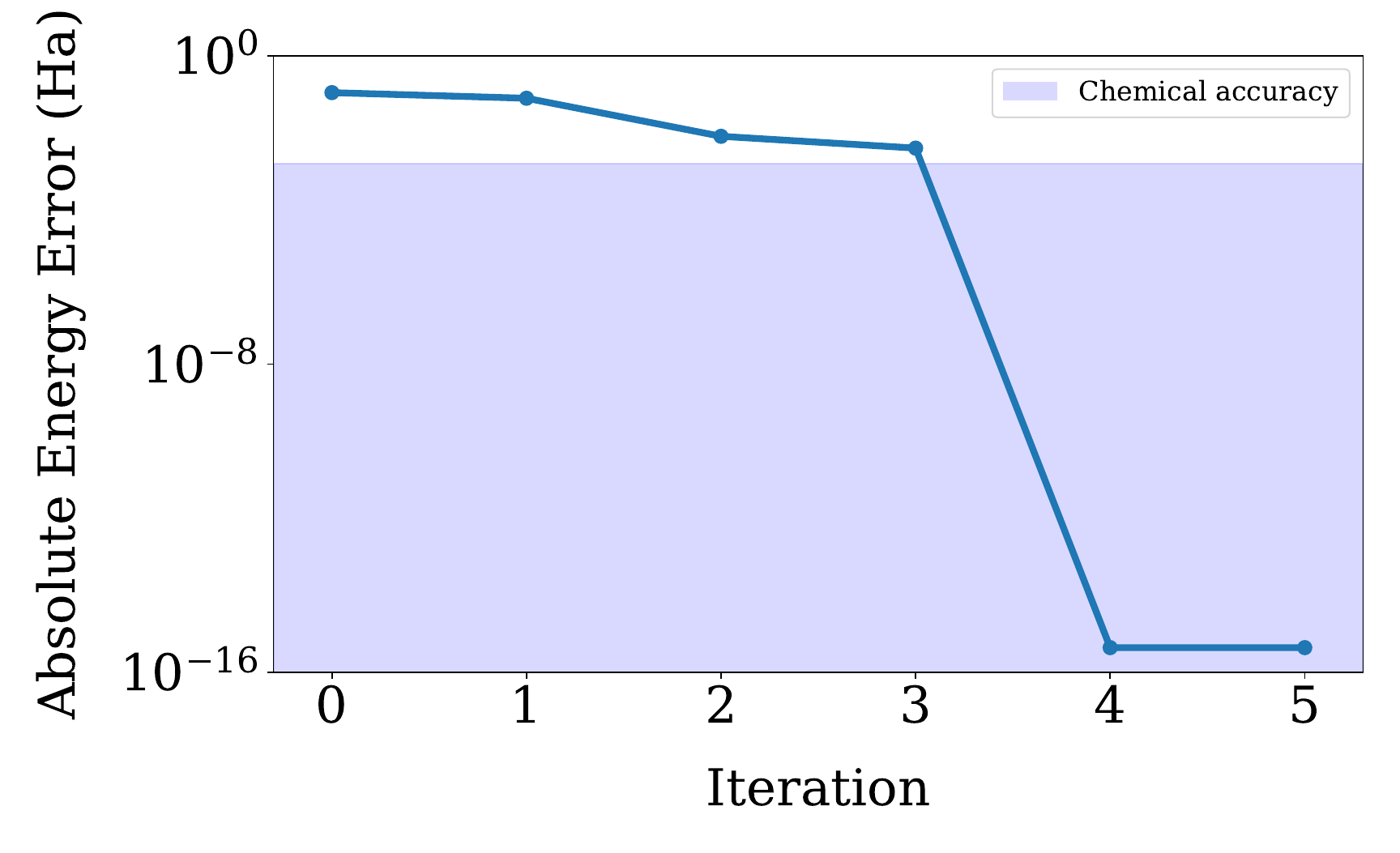}
    \caption{\centering Energy error vs iteration}
    \label{fig:Ideal_Energy}
  \end{subfigure}
  \caption{Gradient heatmap (left) and energy error plot (right) for a noiseless ADAPT-VQE run.}
  \label{fig:ideal_2x1}
\end{figure*}

\begin{figure*}[t]
  \centering

  \begin{subfigure}[t]{0.48\textwidth}
    \centering
    \includegraphics[width=\linewidth]{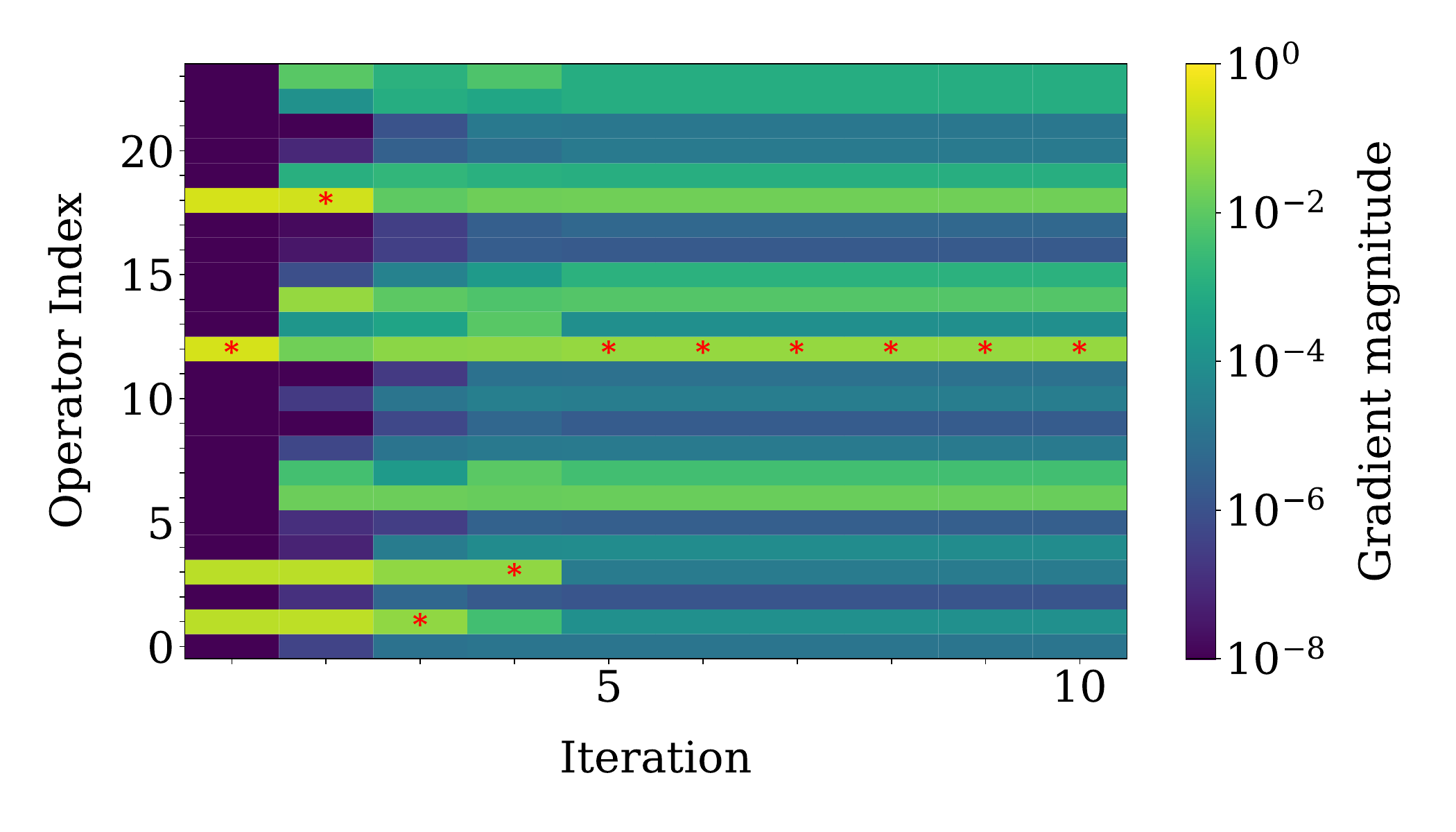}
    \caption{\centering All Incoherent (0.01)}
    \label{fig:hm-001-all}
  \end{subfigure}\hfill
  \begin{subfigure}[t]{0.48\textwidth}
    \centering
    \includegraphics[width=\linewidth]{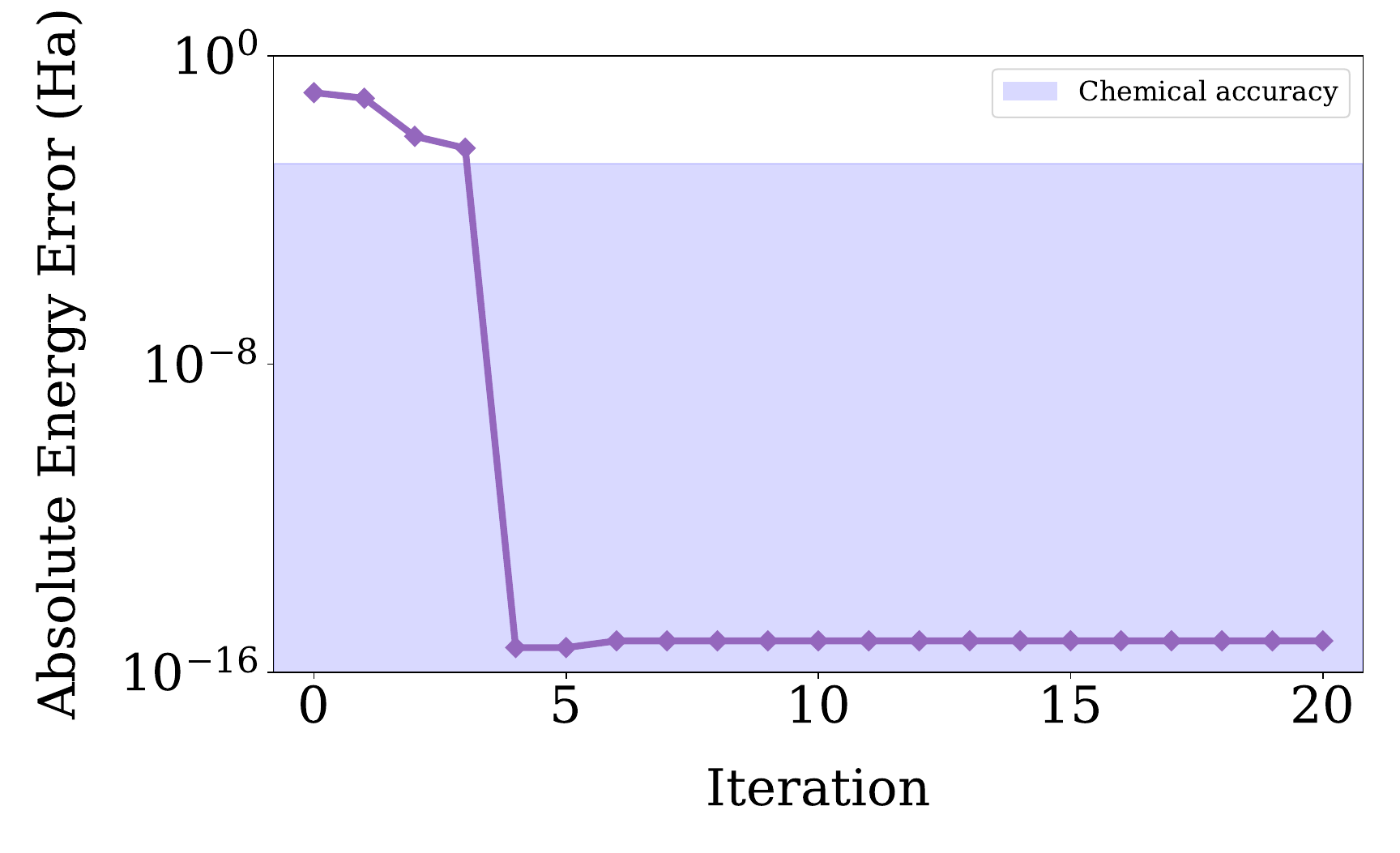}
    \caption{\centering All Incoherent (0.01) }
    \label{fig:ee-001-all}
  \end{subfigure}

  \caption{Gradient heatmap (left) and energy error plots (right) for an ADAPT-VQE run with the `all incoherent' noise channel without error mitigation. The noise strength parameter is set to 0.01.}
  \label{fig:all_incoherent_001}
\end{figure*}

In this section, we show the effect of incoherent noise channels on operator selection and the evolution of the absolute error in the ADAPT-VQE energy. Although our primary focus was operator selection, ultimately, the evolution of the error is what informs us about whether noise had a substantial impact on ADAPT-VQE's performance (and whether error mitigation was necessary and/or successful). Due to similarities among the noise channels and for the sake of brevity, we will discuss only the `all incoherent' noise channel in this section.  The results of non-EM and EM ADAPT-VQE runs for amplitude damping, phase damping, and depolarizing noise can be found in Appendix~\ref{Section VI A}.

The panels in Fig.~\ref{fig:ideal_2x1} show the gradient landscape (\ref{fig:Ideal_Heatmap}) and error convergence (\ref{fig:Ideal_Energy}) in a noiseless setting. In Fig.~\ref{fig:Ideal_Heatmap}, we see that the gradient landscape is quite sparse, with a majority of operators at each iteration having zero gradient. This landscape changes in each iteration, leading to distinct operators being selected in each case. Fig.~\ref{fig:Ideal_Energy} shows that it takes 4 iterations for ADAPT-VQE to converge and terminate, and that it does so well within chemical accuracy.

Fig.~\ref{fig:all_incoherent_001} shows that the injection of noise into the operator selection process significantly impacts the gradient landscape of ADAPT-VQE as it searches for the ground state energy of H$_3$. Noise progressively flattens this landscape beyond the early iterations, causing the gradient magnitudes to stagnate around iteration 5 for this system. This flattening led to repeated selection of the same operator even after the energy error was fully converged. From these results, we conclude that noise is capable of inflating gradient magnitudes well above the convergence threshold, preventing the algorithm from terminating according to the standard convergence criterion even when the state energy is very close to the target, as shown in Fig.~\ref{fig:ee-001-all}. Notably, a noise-aware convergence criterion would be enough to reach a good final ansatz in this case; for example, the algorithm could be stopped when the decrease in energy between the last two iterations falls below a given threshold. This would suffice to reach chemical accuracy, since while this noise level did affect the gradient landscape's characteristics, it did not affect the quality of the selected operators. This confirms that an operator does not need to have the highest energy gradient to induce significant energy changes. Noise at this level still allowed ADAPT-VQE to select effective operators, such that it was still possible to reach a chemically-accurate ground state energy, as shown in Fig.~\ref{fig:ee-001-all}.

Finally, we can also see from these figures that noise may reduce sparsity in the gradient landscape. In the noiseless case (Fig.~\ref{fig:Ideal_Heatmap}), the majority of operators exhibited zero gradients. Under noisy conditions (Fig.~\ref{fig:hm-001-all}), all four noise channels yielded substantially more non-zero gradients after the first iteration.

\begin{figure*}[t]
  \centering

  \makebox[\textwidth][c]{%
    \begin{subfigure}[t]{0.45\textwidth}
      \centering
      \includegraphics[width=\linewidth]{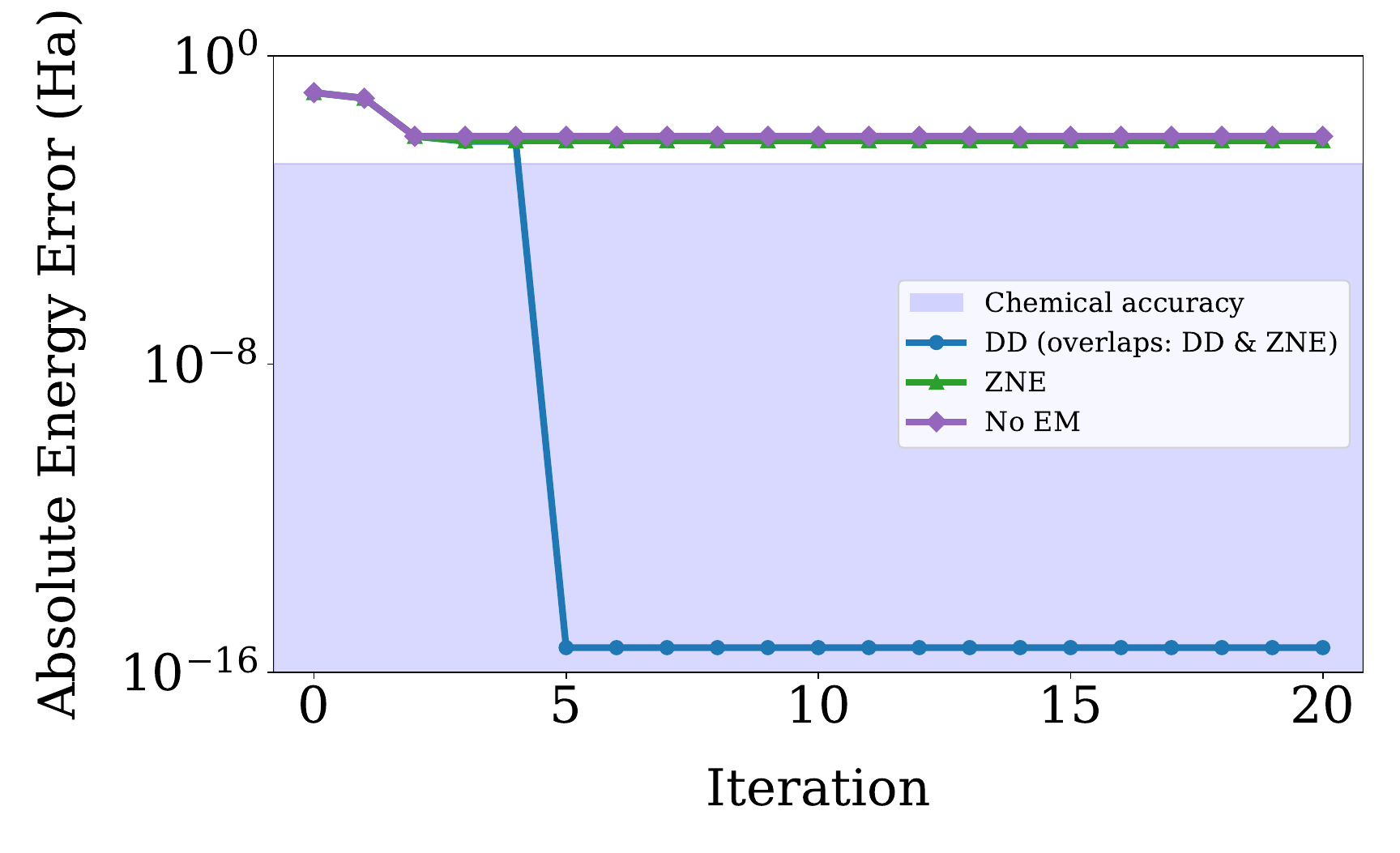}
      \caption{\centering All Incoherent (0.1)}
      \label{fig:ee-allinco-01}
    \end{subfigure}
  }


  \begin{subfigure}[t]{0.45\textwidth}
    \centering
    \includegraphics[width=\linewidth]{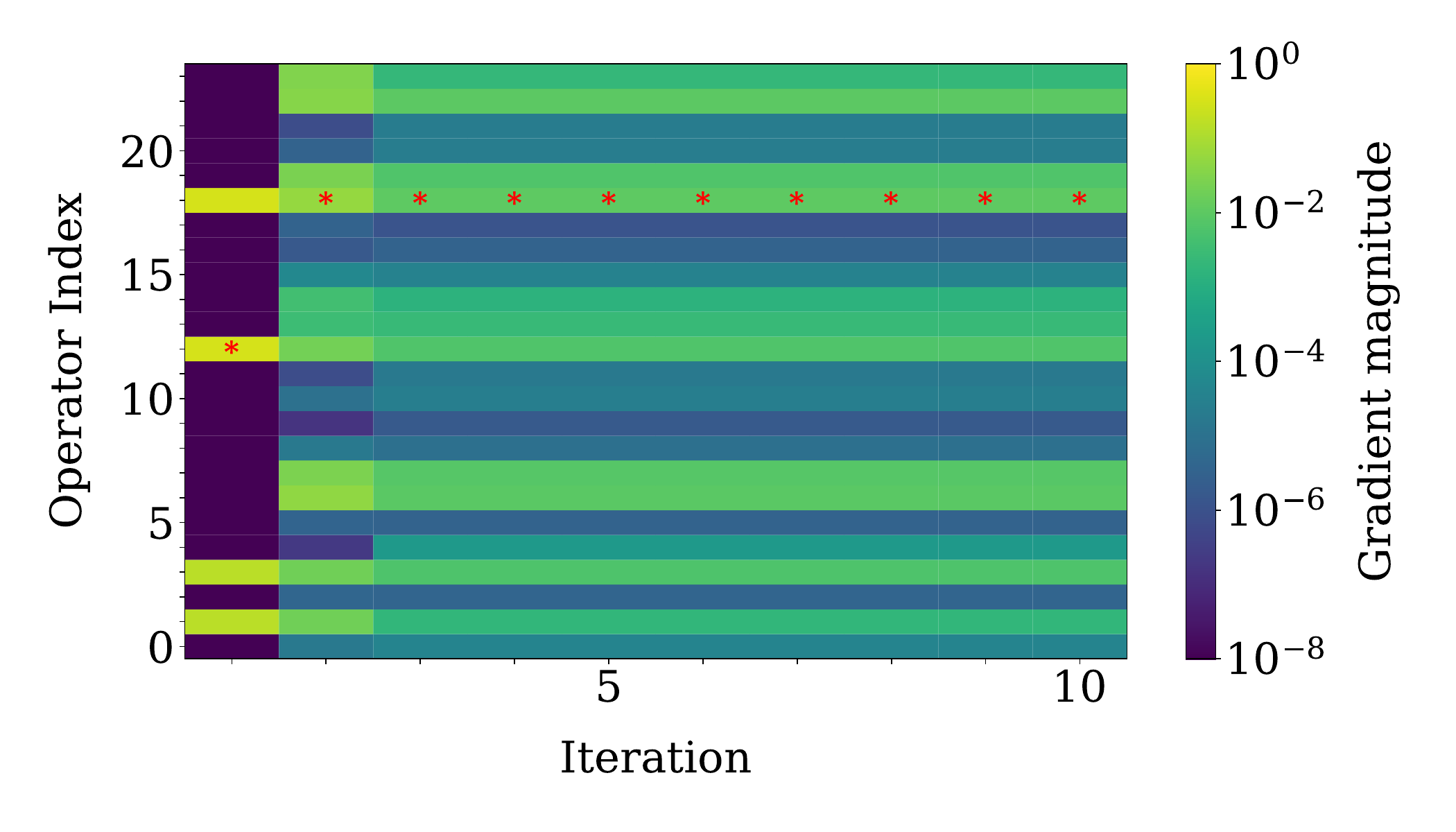}
    \caption{\centering No EM (0.1)}
    \label{fig:hm-allinco-01-noEM}
  \end{subfigure}\hfill
  \begin{subfigure}[t]{0.45\textwidth}
    \centering
    \includegraphics[width=\linewidth]{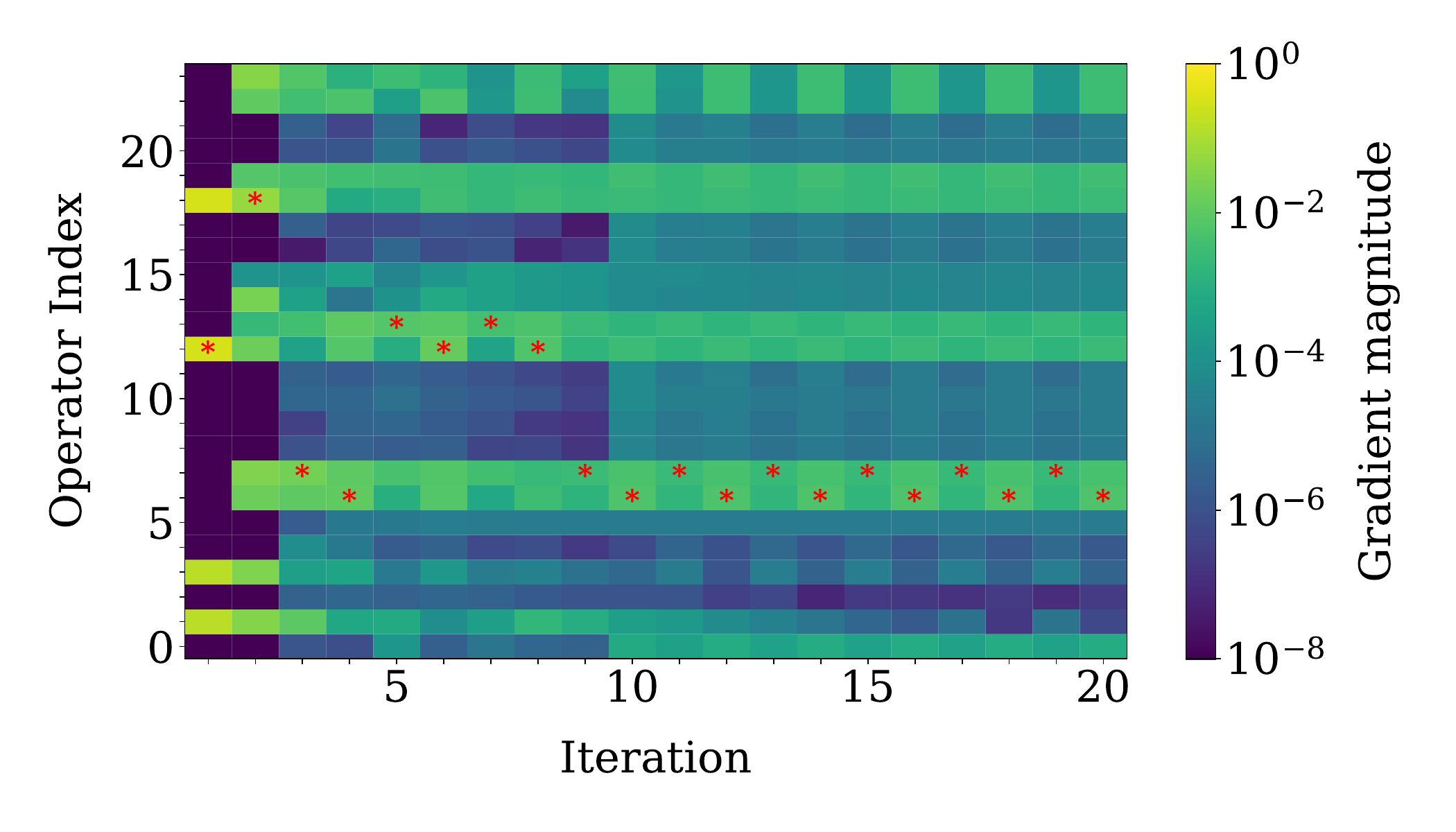}
    \caption{\centering DD (0.1)}
    \label{fig:hm-allinco-01-dd}
  \end{subfigure}


  \begin{subfigure}[t]{0.45\textwidth}
    \centering
    \includegraphics[width=\linewidth]{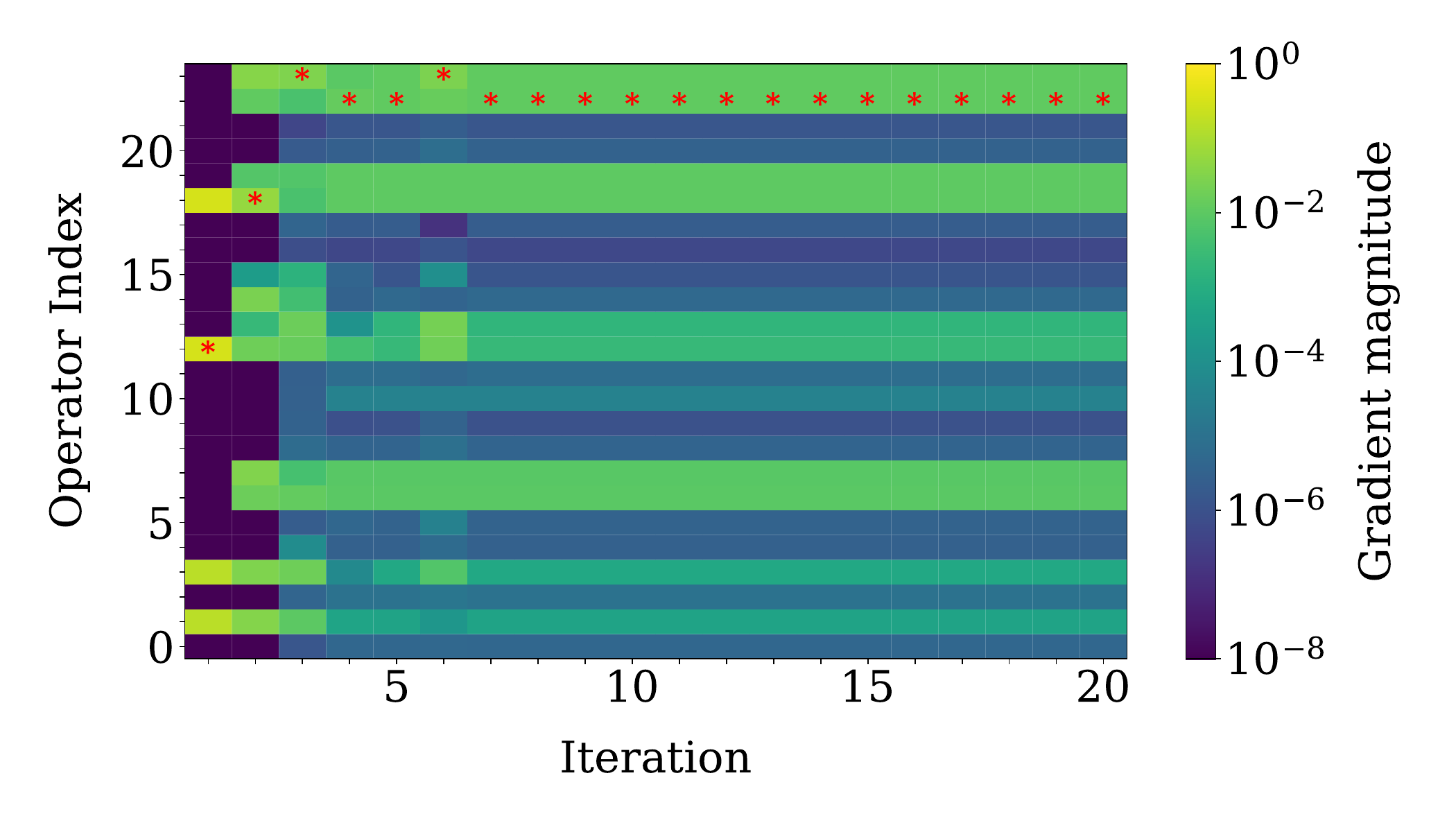}
    \caption{\centering ZNE (0.1)}
    \label{fig:hm-allinco-01-zne}
  \end{subfigure}\hfill
  \begin{subfigure}[t]{0.45\textwidth}
    \centering
    \includegraphics[width=\linewidth]{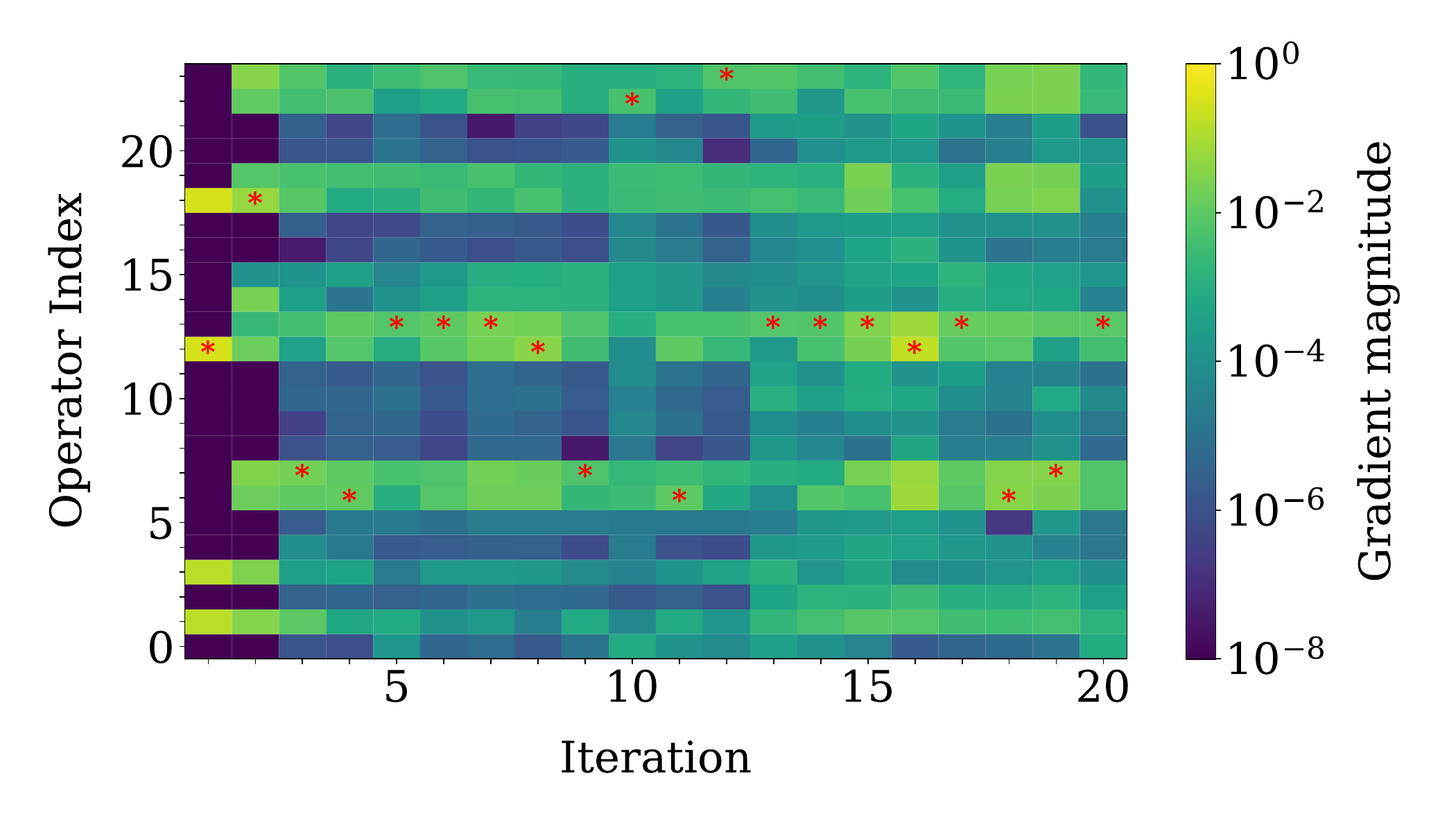}
    \caption{\centering DD \& ZNE (0.1)}
    \label{fig:hm-allinco-01-ddzne}
  \end{subfigure}

  \caption{`All incoherent' noise channel results at strength 0.1: energy-error plots (a), and gradient heatmaps (b-e) under different EM settings.}
  \label{fig:hm-01-allinco}
\end{figure*}
\begin{figure*}[t]
  \centering

  \makebox[\textwidth][c]{%
    \begin{subfigure}[t]{0.45\textwidth}
      \centering
      \includegraphics[width=\linewidth]{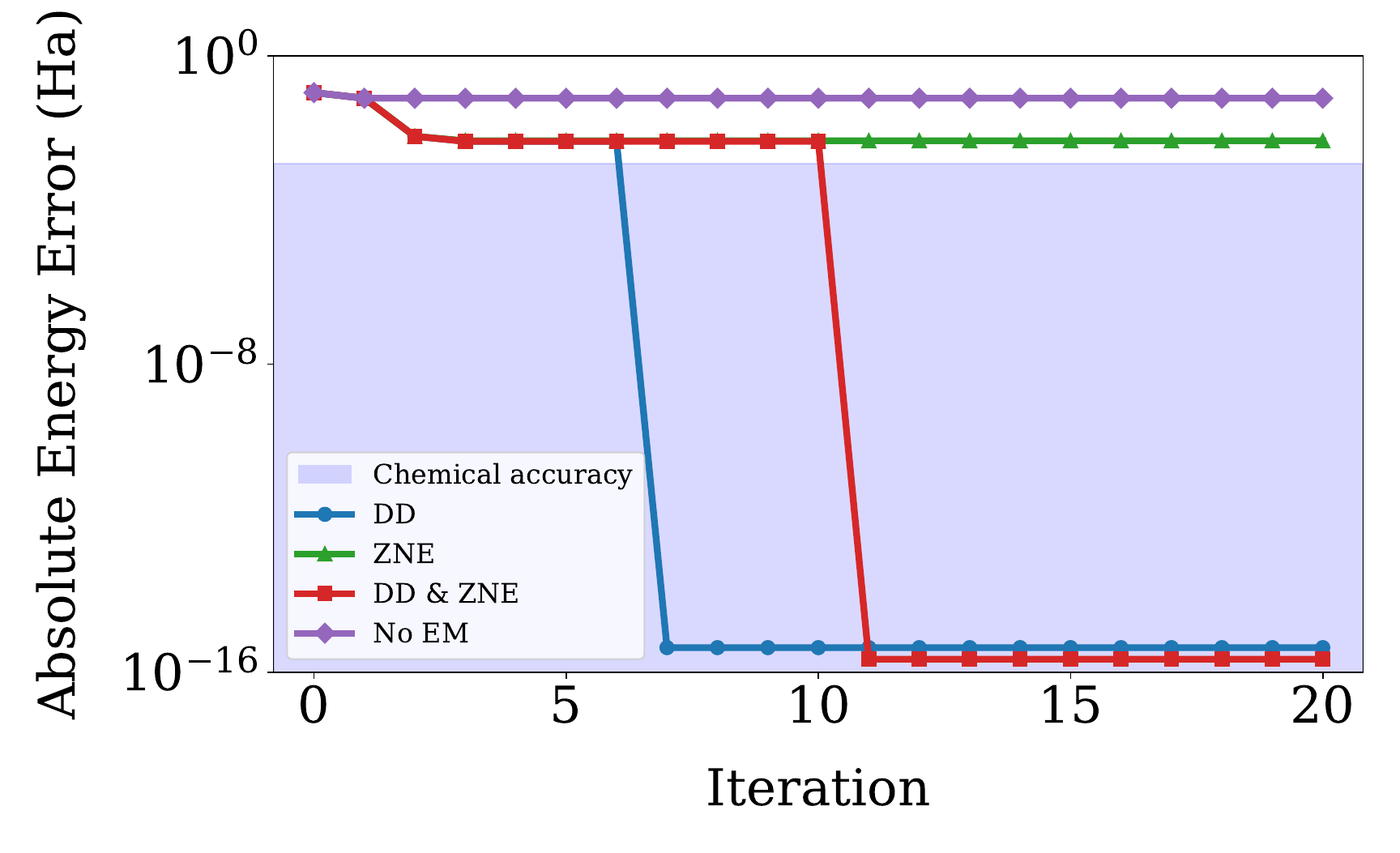}
      \caption{\centering All Incoherent (0.3)}
      \label{fig:ee-allinco-03}
    \end{subfigure}
  }


  \begin{subfigure}[t]{0.45\textwidth}
    \centering
    \includegraphics[width=\linewidth]{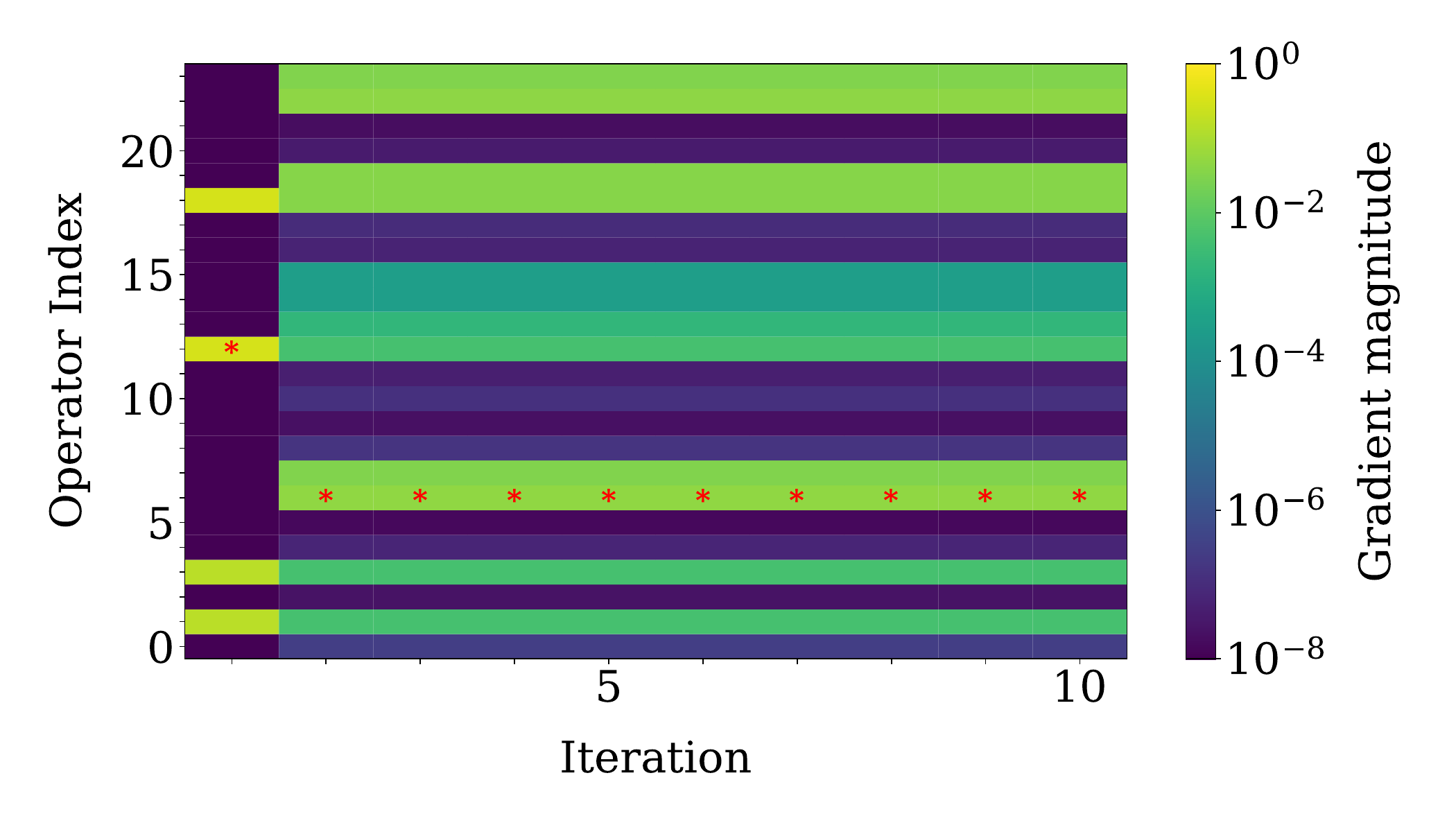}
    \caption{\centering No EM (0.3)}
    \label{fig:hm-03-allinco-noEM}
  \end{subfigure}\hfill
  \begin{subfigure}[t]{0.45\textwidth}
    \centering
    \includegraphics[width=\linewidth]{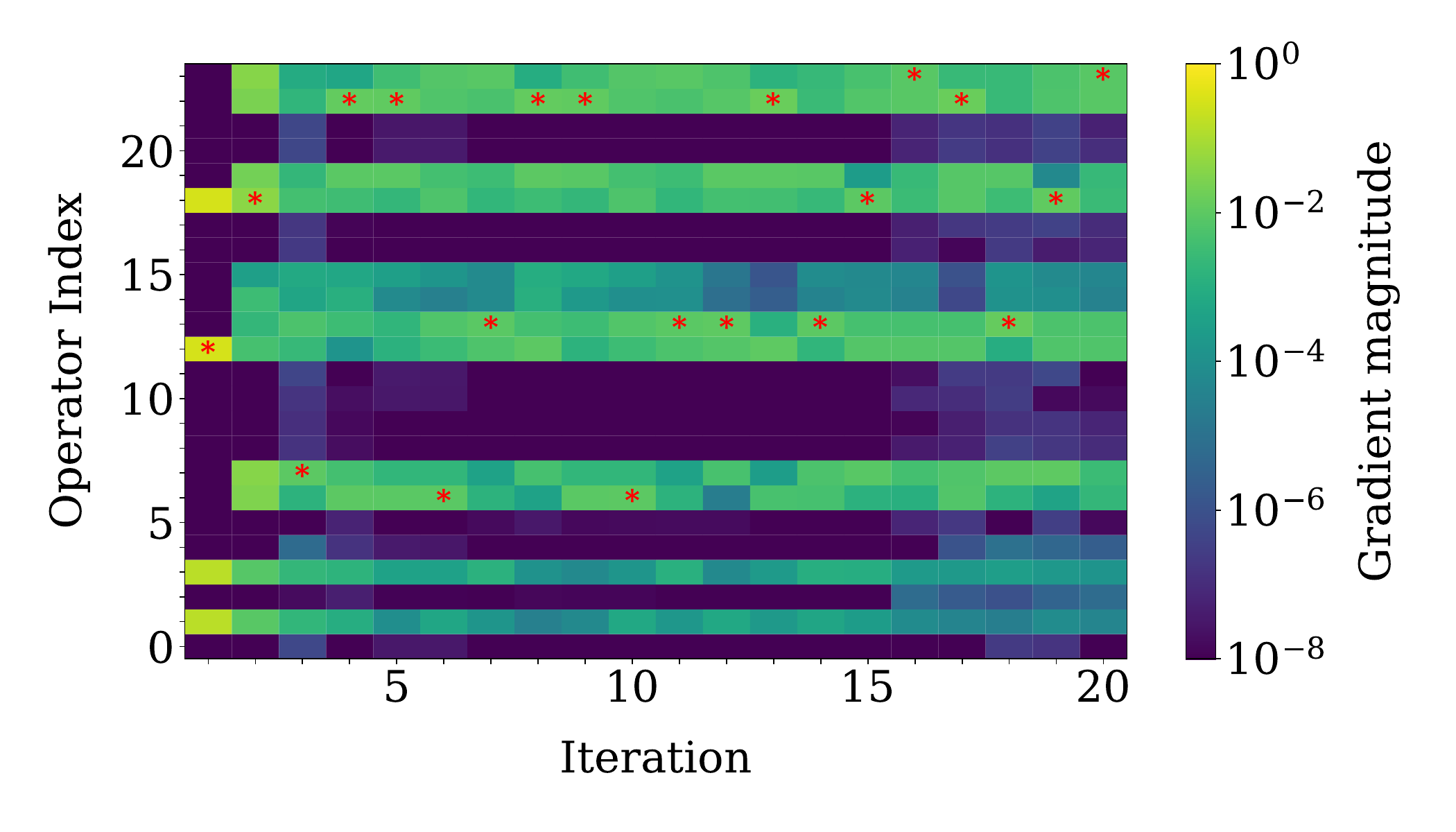}
    \caption{\centering DD (0.3)}
    \label{fig:hm-03-allinco-dd}
  \end{subfigure}

  \vspace{0.5em}

  \begin{subfigure}[t]{0.45\textwidth}
    \centering
    \includegraphics[width=\linewidth]{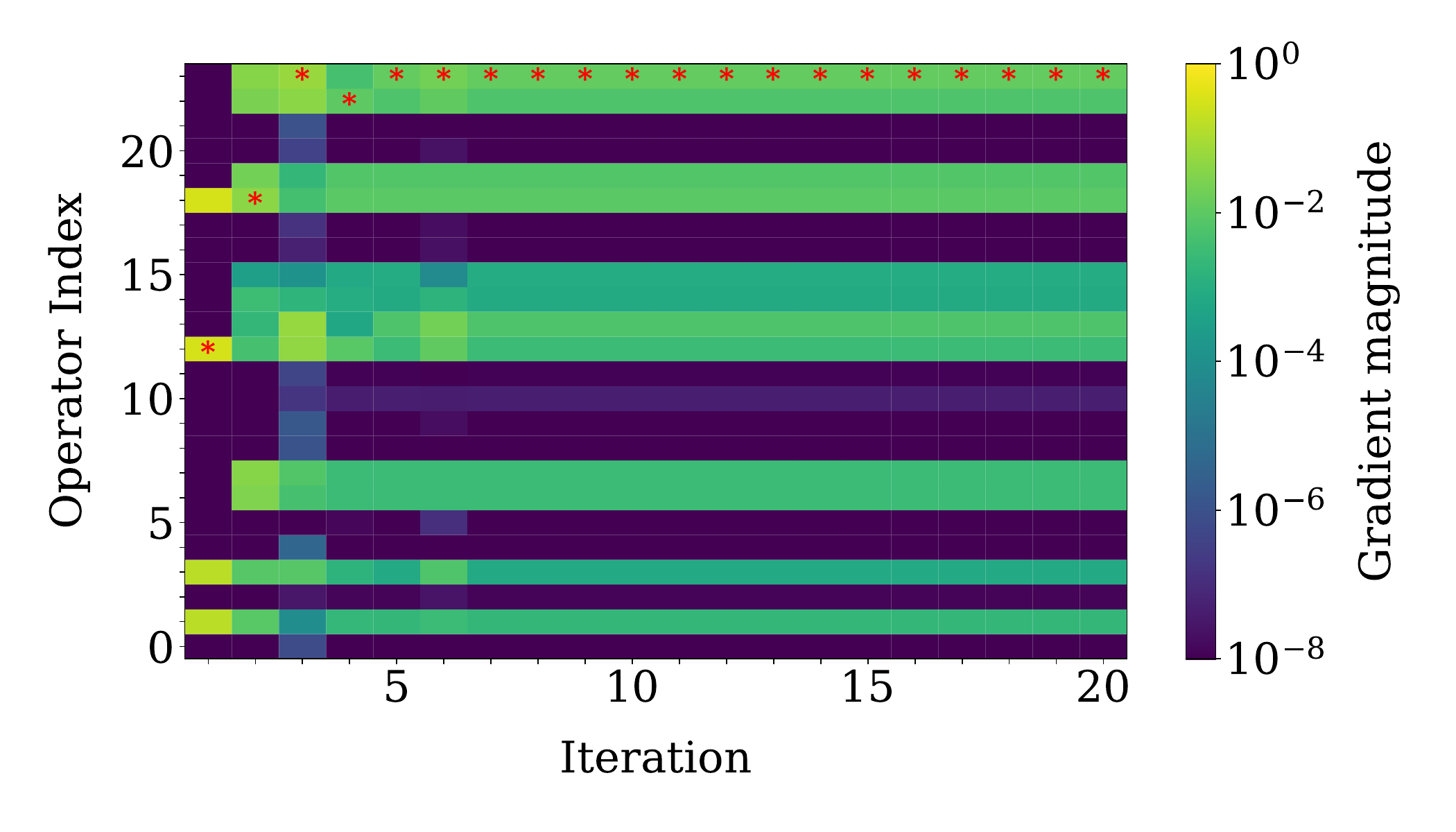}
    \caption{\centering ZNE (0.3)}
    \label{fig:hm-03-allinco-zne}
  \end{subfigure}\hfill
  \begin{subfigure}[t]{0.45\textwidth}
    \centering
    \includegraphics[width=\linewidth]{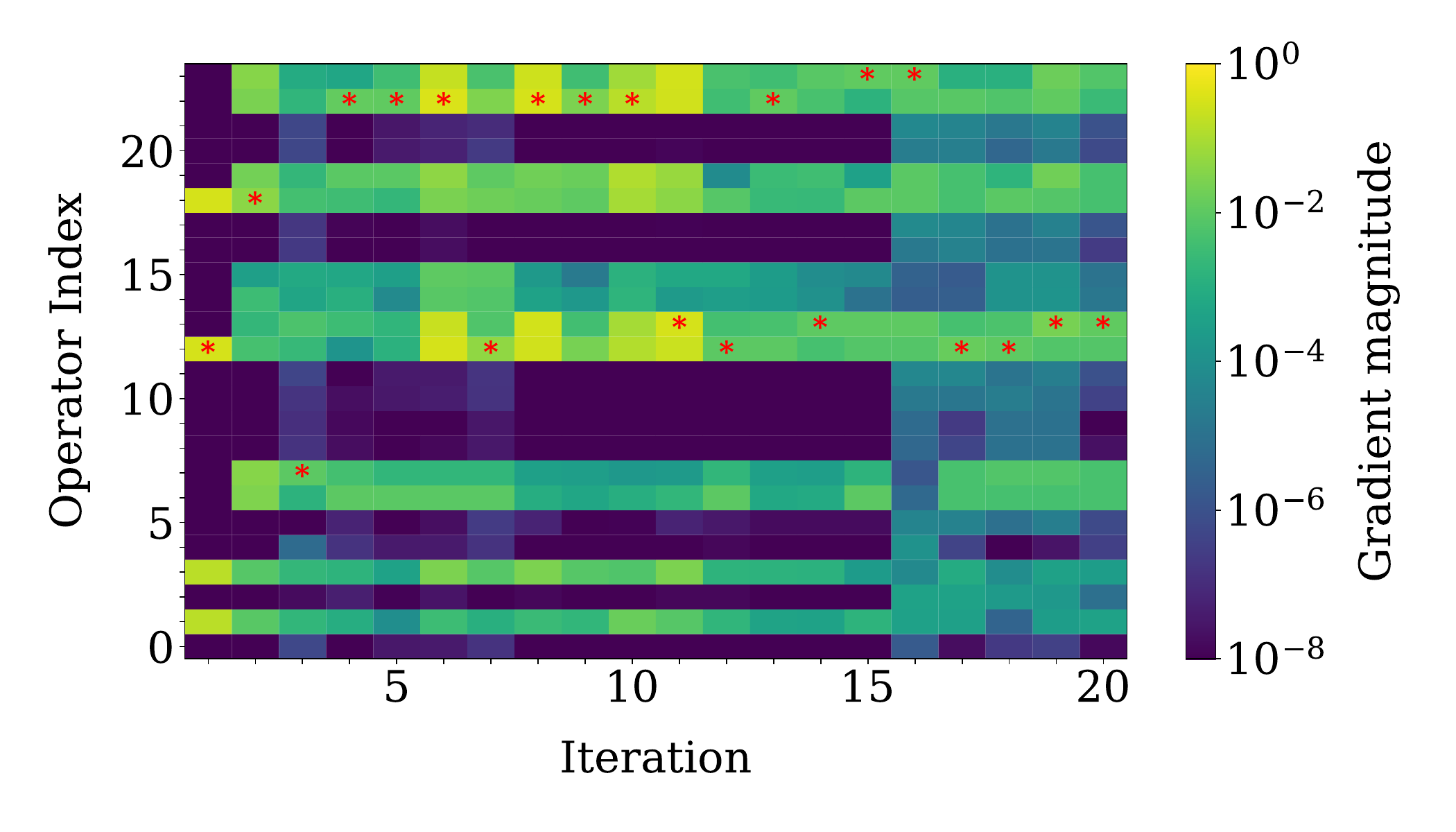}
    \caption{\centering DD \& ZNE (0.3)}
    \label{fig:hm-03-allinco-ddzne}
  \end{subfigure}

  \caption{`All incoherent' noise channel results at strength $0.3$: energy-error plots (a), and gradient heatmaps (b-e) under different EM settings.}
  \label{fig:hm-03-allinco}
\end{figure*}


Evidently, a higher noise strength will have a higher impact on ADAPT-VQE performance. For the noise strength of Fig.~\ref{fig:all_incoherent_001}, none of the four noise models impacted the energy error significantly. We now consider the impact of higher noise levels. Figures~\ref{fig:ee-allinco-01} and \ref{fig:ee-allinco-03} show the absolute energy error in the case where the noise parameter is set to 0.1 and 0.3, respectively. In both cases, the all-incoherent noise channel without EM resulted in a significantly higher energy error, preventing chemical accuracy from being reached.

The gradient landscape also changed in response to differing noise levels. We observed a trend for the gradient landscape to stagnate after an increasingly shorter number of iterations: For the lowest noise strength (Fig~.\ref{fig:ee-001-all}), this happened after five iterations, while for the stronger noise channels, as shown by Figs.~\ref{fig:hm-allinco-01-noEM}~\ref{fig:hm-03-allinco-noEM},  this stagnation happened after two iterations.

From Figs.~\ref{fig:ee-allinco-01} and \ref{fig:ee-allinco-03}, we can see the impact of EM on the convergence of the ADAPT-VQE error. We observe that, in many cases, EM effectively mitigates the impact of noise on operator selection. At the 0.1 and 0.3 noise levels, noise prevents ADAPT-VQE from reaching chemical accuracy by leading to poor operator selection. However, error mitigation strategies are effective at restoring chemical accuracy. Used separately, DD and ZNE were only successful in restoring chemical accuracy in some cases. However, when used together, they enabled the algorithm to reach chemical accuracy in all cases we tested.

\begin{figure*}[t]
  \centering

  \makebox[\textwidth][c]{%
    \begin{subfigure}[t]{0.45\textwidth}
      \centering
      \includegraphics[width=\linewidth]{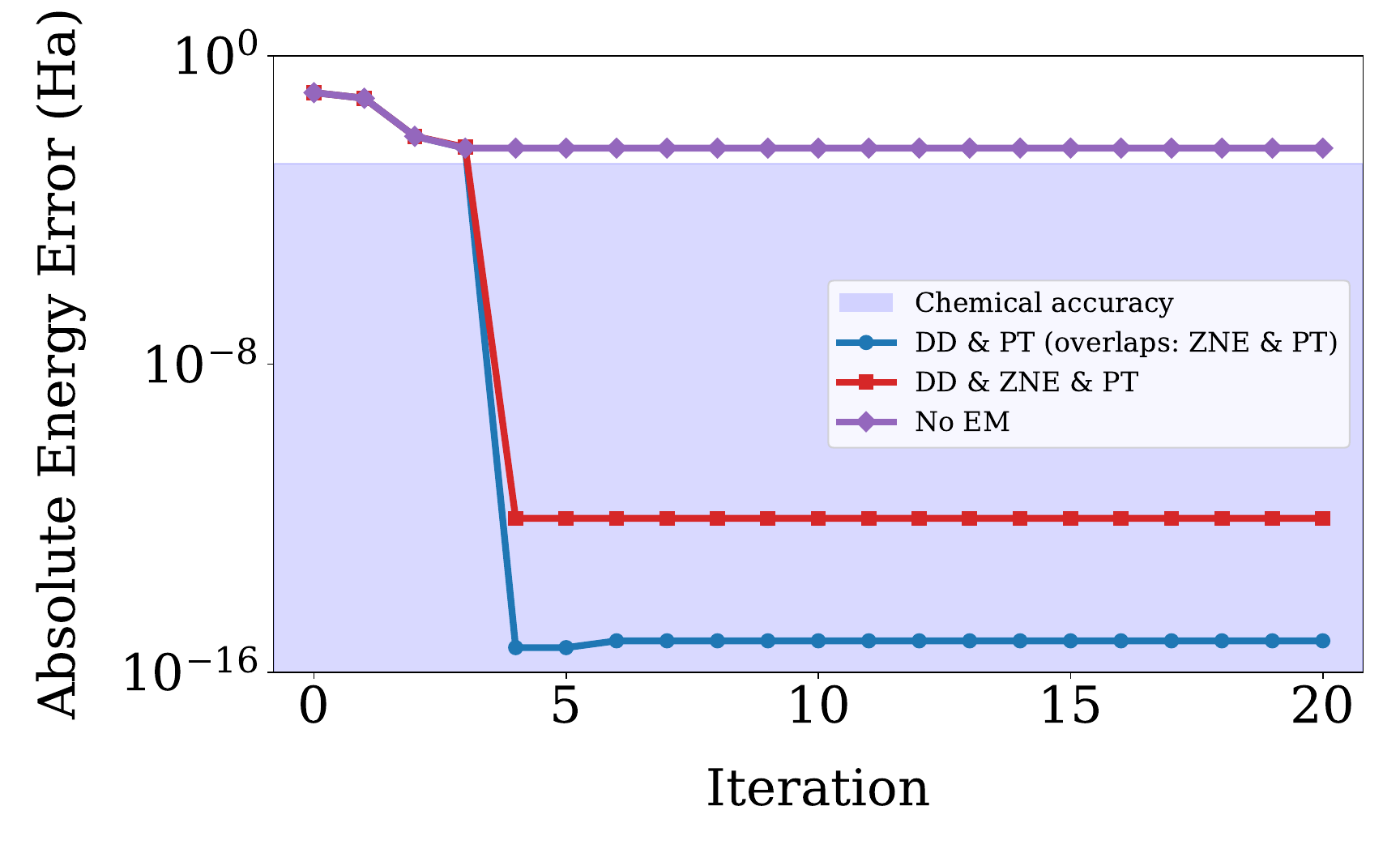}
      \caption{\centering All Coherent (0.1)}
      \label{fig:ee-allcoh-01}
    \end{subfigure}
  }


  \begin{subfigure}[t]{0.45\textwidth}
    \centering
    \includegraphics[width=\linewidth]{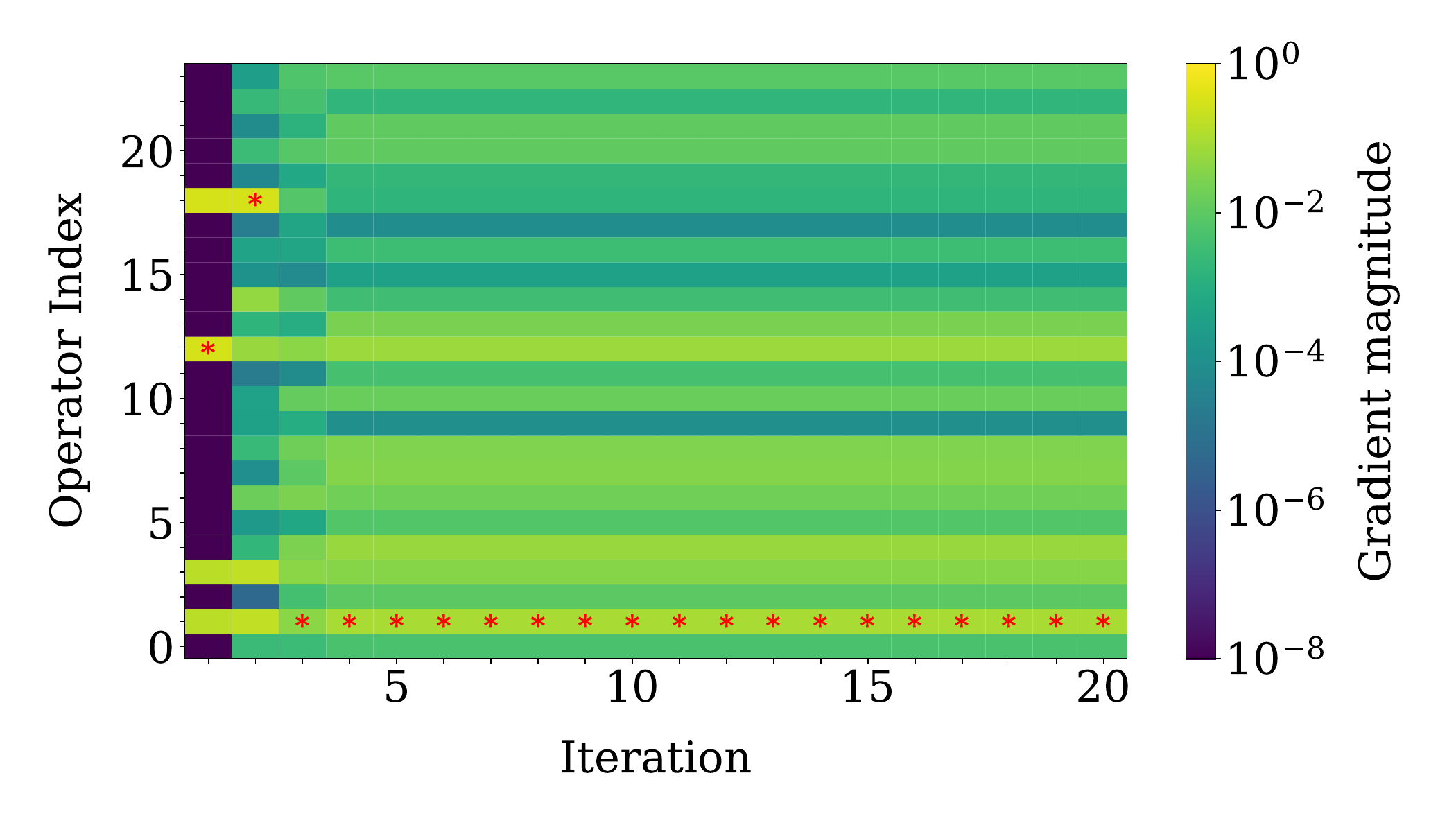}
    \caption{\centering No EM (0.1)}
    \label{fig:hm-01-allcoh-noEM}
  \end{subfigure}\hfill
  \begin{subfigure}[t]{0.45\textwidth}
    \centering
    \includegraphics[width=\linewidth]{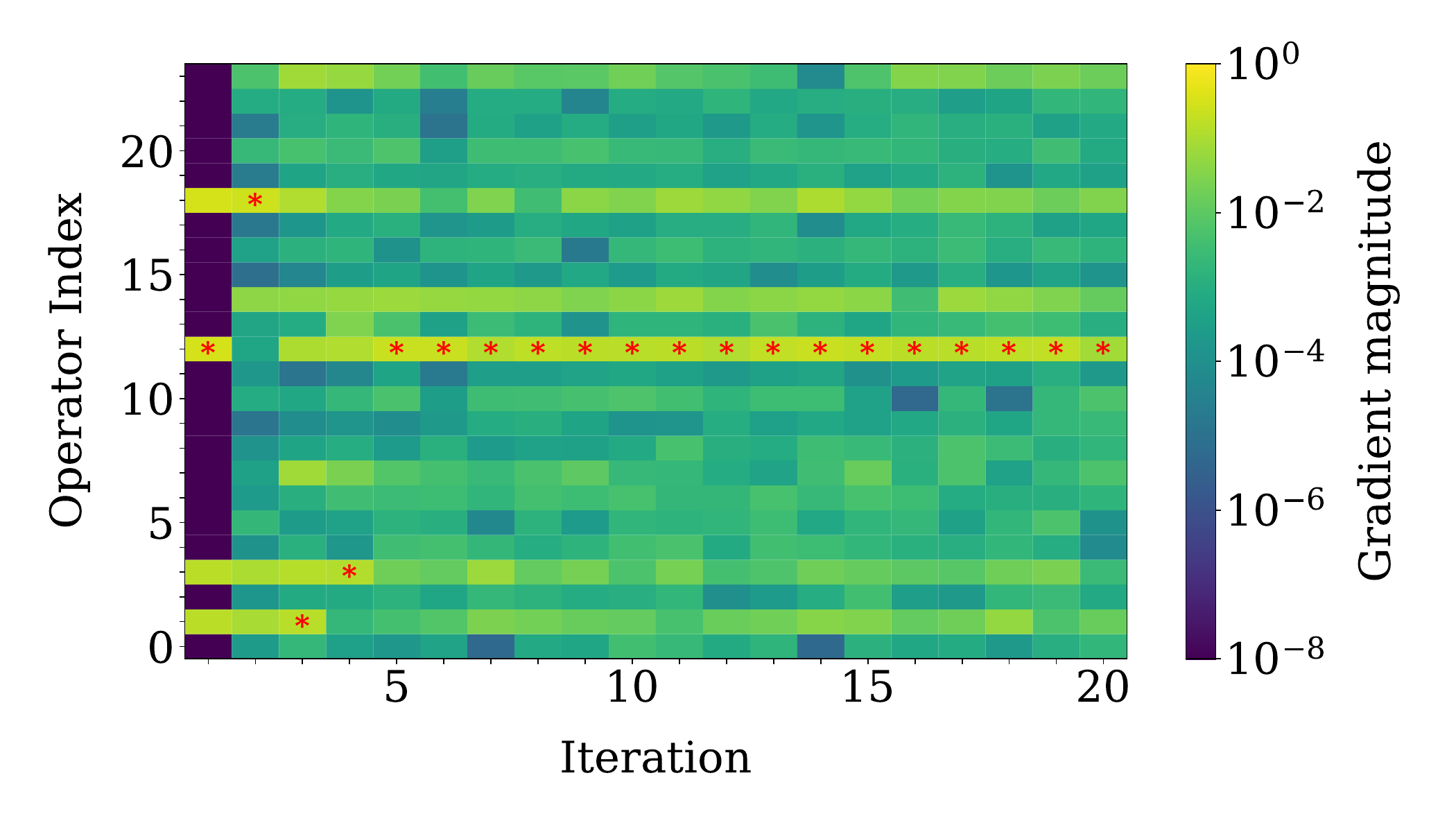}
    \caption{\centering DD \& PT (0.1)}
    \label{fig:hm-01-allcoh-ddp}
  \end{subfigure}


  \begin{subfigure}[t]{0.45\textwidth}
    \centering
    \includegraphics[width=\linewidth]{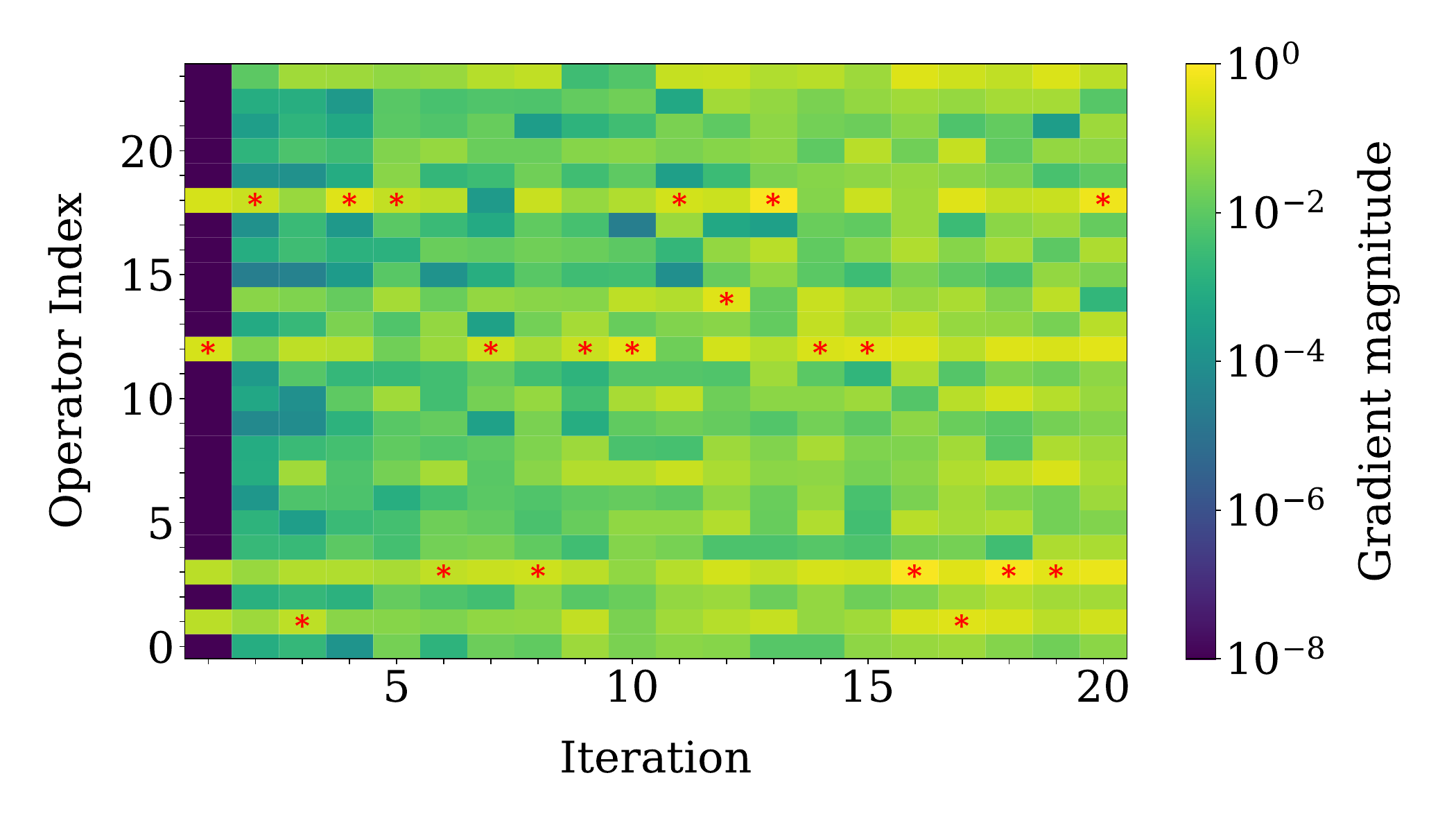}
    \caption{\centering ZNE \& PT (0.1)}
    \label{fig:hm-01-allcoh-pzne}
  \end{subfigure}\hfill
  \begin{subfigure}[t]{0.45\textwidth}
    \centering
    \includegraphics[width=\linewidth]{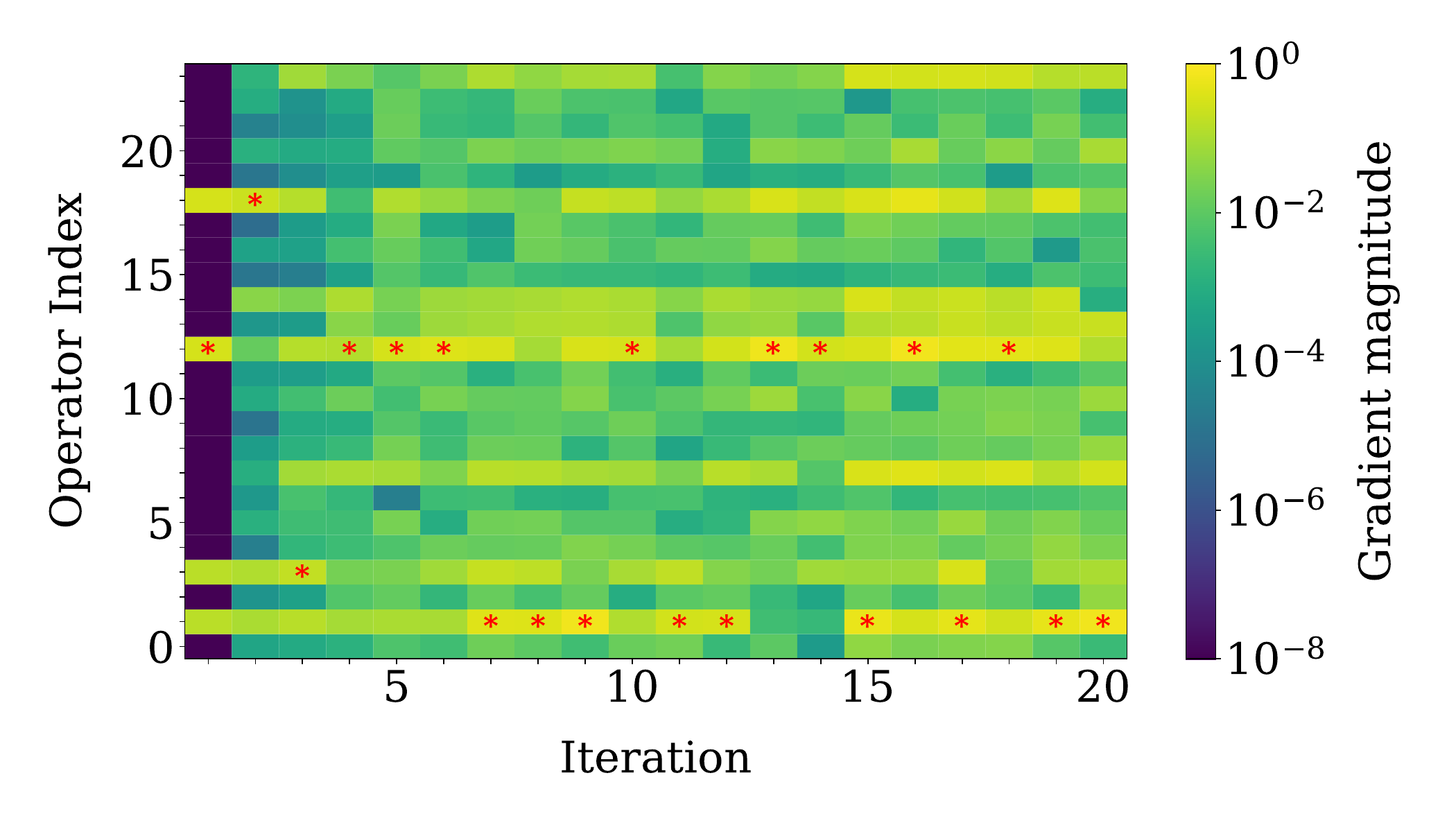}
    \caption{\centering DD \& ZNE \& PT (0.1)}
    \label{fig:hm-01-allcoh-ddpzne}
  \end{subfigure}

  \caption{`All coherent' noise results at strength $0.1$: energy-error plots (a), and gradient heatmaps (b-e) under different EM settings.}
  \label{fig:allcoh_01_combined}
\end{figure*}

\begin{figure*}[t]
  \centering

  \makebox[\textwidth][c]{%
    \begin{subfigure}[t]{0.45\textwidth}
      \centering
      \includegraphics[width=\linewidth]{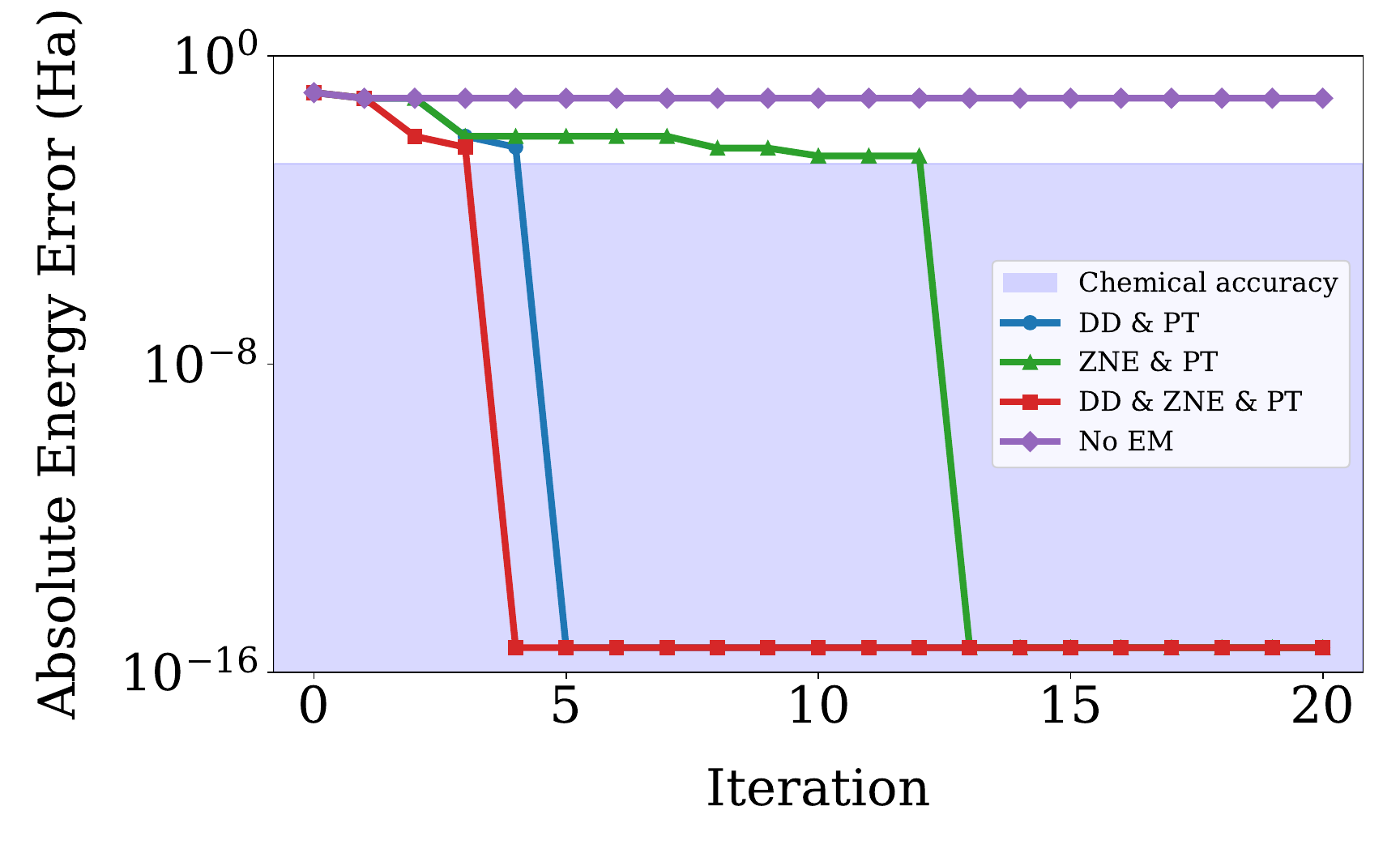}
      \caption{\centering All Coherent (0.3)}
      \label{fig:ee-allcoh-03}
    \end{subfigure}
  }


  \begin{subfigure}[t]{0.45\textwidth}
    \centering
    \includegraphics[width=\linewidth]{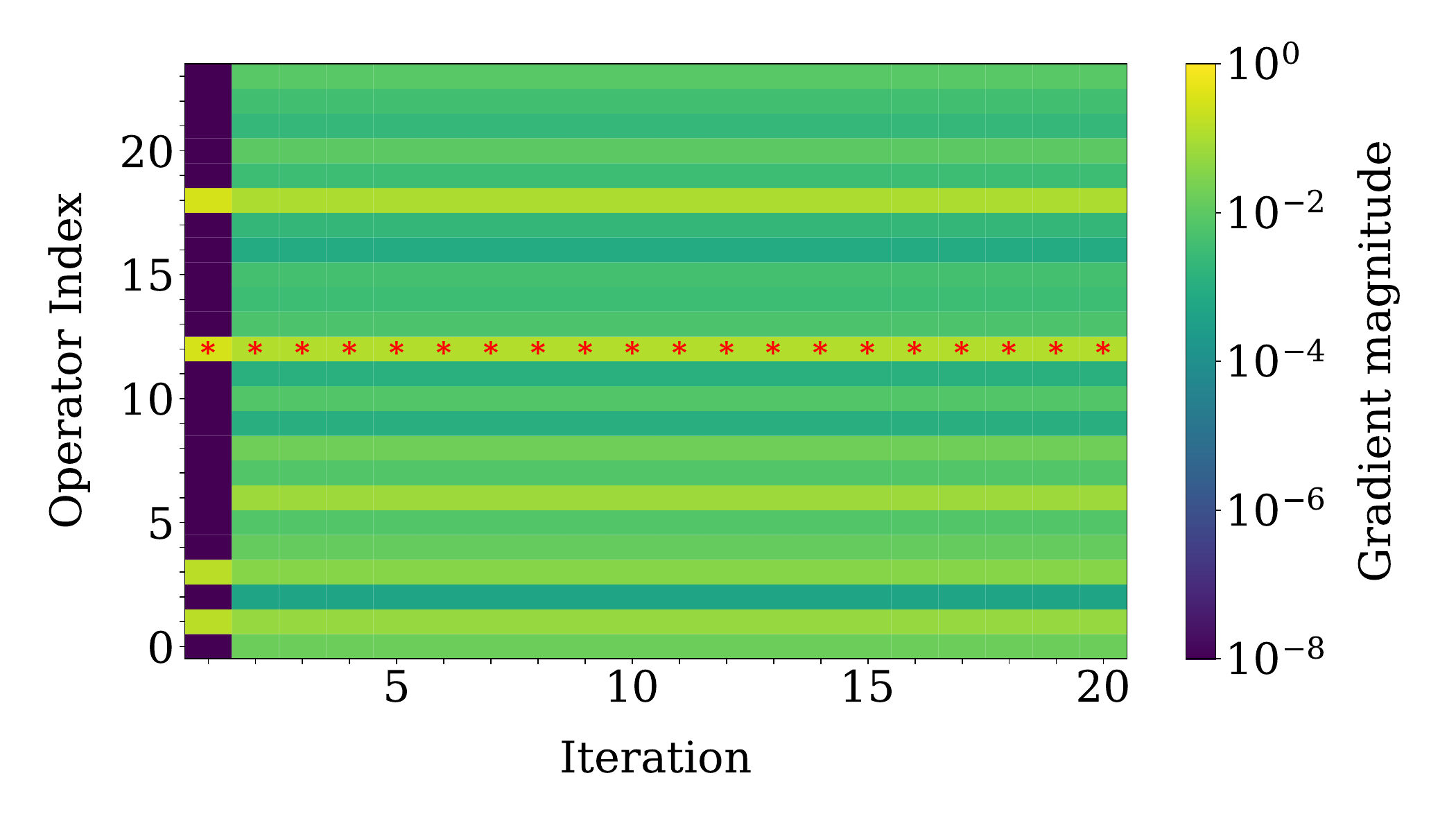}
    \caption{\centering No EM (0.3)}
    \label{fig:hm-03-allcoh-noEM}
  \end{subfigure}\hfill
  \begin{subfigure}[t]{0.45\textwidth}
    \centering
    \includegraphics[width=\linewidth]{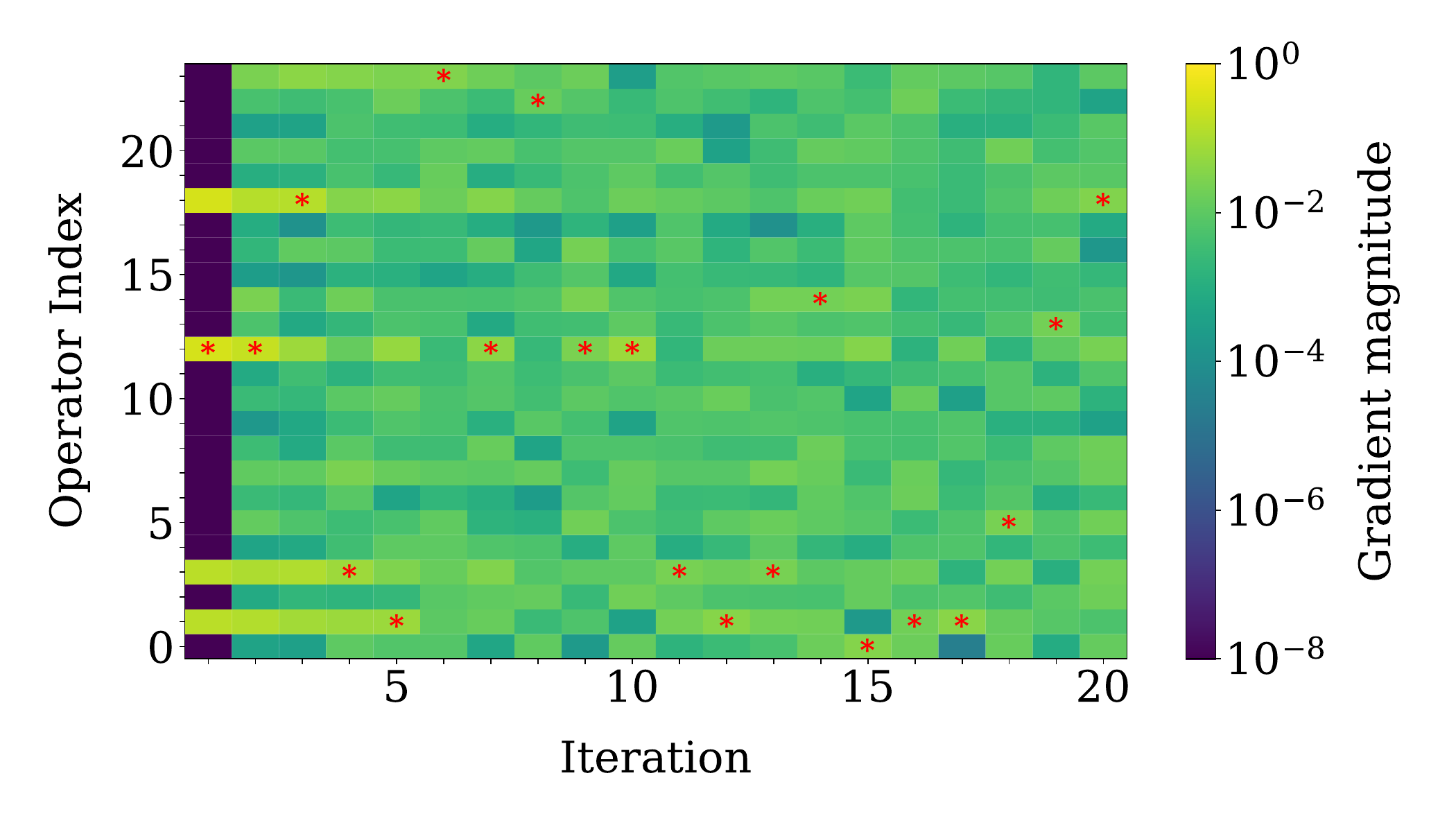}
    \caption{\centering DD \& PT (0.3)}
    \label{fig:hm-03-allcoh-ddp}
  \end{subfigure}


  \begin{subfigure}[t]{0.45\textwidth}
    \centering
    \includegraphics[width=\linewidth]{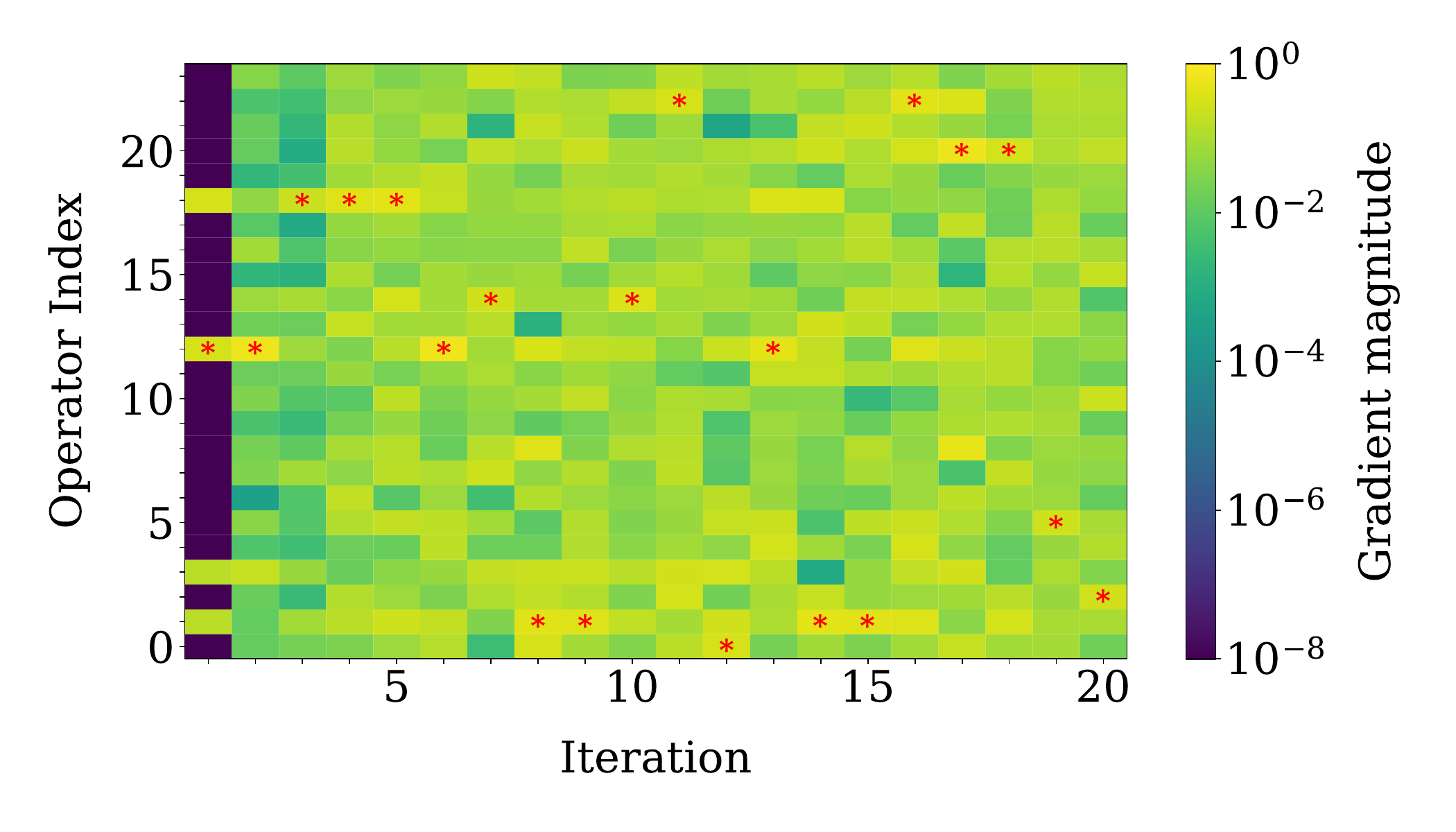}
    \caption{\centering ZNE \& PT (0.3)}
    \label{fig:hm-03-allcoh-pzne}
  \end{subfigure}\hfill
  \begin{subfigure}[t]{0.45\textwidth}
    \centering
    \includegraphics[width=\linewidth]{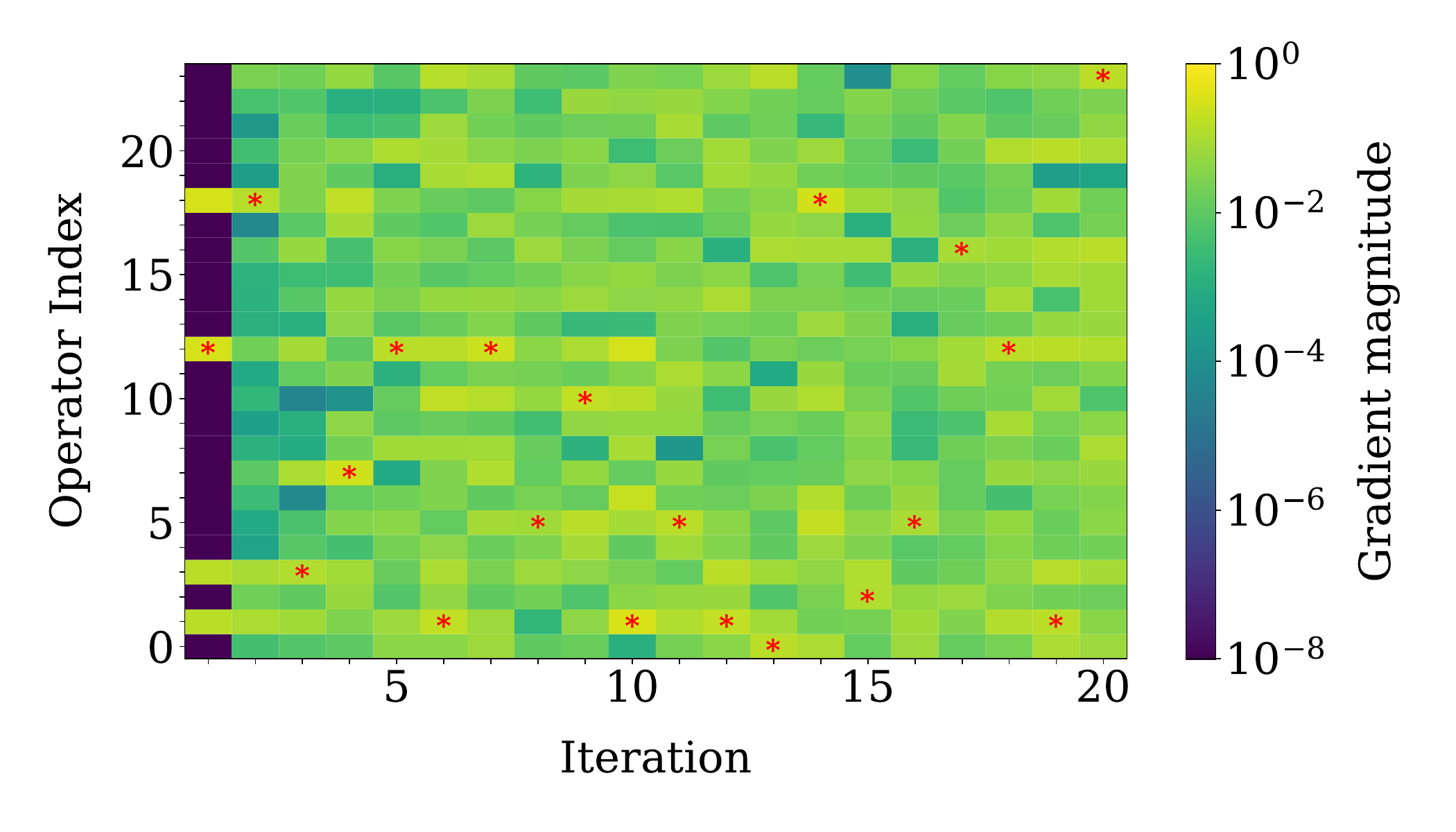}
    \caption{\centering DD \& ZNE \& PT (0.3)}
    \label{fig:hm-03-allcoh-ddpzne}
  \end{subfigure}

  \caption{`All coherent' noise channel results at strength $0.3$: energy-error plots (a), and gradient heatmaps (b-e) under different EM settings.}
  \label{fig:allcoh_03_combined}
\end{figure*}

Figures~\ref{fig:hm-01-allinco} and \ref{fig:hm-03-allinco} additionally show the impact of EM on the gradient landscapes. We observe that in all cases, error mitigation strategies either delay or prevent the gradient landscape from stagnating. There is an interesting relationship between whether the stagnation of the gradient landscape was prevented or delayed and whether chemical accuracy was reached in 20 iterations: For runs in which the gradient landscape's stagnation was postponed, chemical accuracy was not reached; for runs in which stagnation was prevented altogether, chemical accuracy was reached. Figs.~\ref{fig:ee-allinco-03} (evolution of energy error for ADAPT runs with various EM techniques),  \ref{fig:hm-03-allinco-dd} (gradient landscape for a run with DD), and \ref{fig:hm-03-allinco-zne} (gradient landscape for a run with ZNE) illustrate this trend. Hence, preventing the stagnation of the gradient landscape can be as a significant factor in achieving chemical accuracy.

We note that EM did not completely restore operator selection to the noiseless case; even when error mitigation methods did restore chemical accuracy, they did not replicate the noiseless ansatz and did not allow ADAPT-VQE to converge and terminate according to the standard criterion. Nevertheless, this was not a limitation of error mitigation because we do not need the selected operator to be exactly the same as in the noiseless case. Several operators produce significant energy changes. Hence, as long as chemical accuracy is reached with the algorithm, we consider EM to have been successful.

\subsection{Coherent Noise} \label{Section IV B}

The purpose of this subsection is to analyze the effect of coherent noise on operator selection and understand whether or not error mitigation strategies can restore chemical accuracy and proper operator selection in such a setting. As before, for the sake of brevity, only gradient heatmaps for the `all coherent' channels will be shown.  Gradient heatmaps for the $X$ and $Z$ over-rotation channels without EM can be found in Appendix~\ref{Section VI B}.

We observe that coherent noise has very similar effects to incoherent noise, as it also leads to stagnated gradients and prevents the convergence of the ADAPT-VQE algorithm. However, coherent noise generally differs from incoherent noise in that it leads to a greater inflation of gradient values. Coherent noise results in \textit{all} operators having non-zero gradient values. In Figs.~\ref{fig:allcoh_01_combined} and \ref{fig:allcoh_03_combined}, for example, we can see that `all coherent' noise channel heatmaps are significantly lighter in color than their incoherent counterparts (Figs.~\ref{fig:hm-01-allinco} and \ref{fig:hm-03-allinco}), indicating that the coherent noise channels have inflated gradient values to a much greater extent. In fact, what we found was that coherent noise leaves all operator gradients in the pool with non-zero gradients across the entirety of the run.\ref{fig:hm-01-allcoh-noEM} and \ref{fig:hm-03-allcoh-noEM} show the impact of coherent noise on the gradient landscape for noise strength 0.1

In Figs.~\ref{fig:ee-allcoh-01} and ~\ref{fig:ee-allcoh-03}, we observe the performance of ADAPT-VQE with and without error mitigation. We observed that the weakest channels (0.01) have no meaningful impact on the final accuracy, so they were omitted. However, for the settings with higher noise rates, coherent noise prevents ADAPT-VQE from reaching chemical accuracy. 

Error mitigation methods were found to be highly effective in reversing the effects of coherent noise and restoring chemical accuracy. In fact, as we can see in Fig. (\ref{fig:ee-allcoh-01}), all error mitigation strategies are able to reach chemical accuracy before the 20th iteration. Previously, we noted that different studies offered contrasting views on whether PT was more effective when paired with DD or ZNE. 
In our case study, however, both pairings proved beneficial, with the combination of all three techniques yielding strong results as well.

The heatmaps in Figures.~\ref{fig:allcoh_01_combined} and \ref{fig:allcoh_03_combined} show the impact of EM on the gradient landscapes. The effects of error mitigation with coherent noise were similar to those of incoherent noise. These heatmaps show that error mitigation strategies prevent the gradient landscape from stagnating in the first 20 iterations, allowing for a different operator to be selected at each iteration. The effects of error mitigation with coherent noise did, however, differ from incoherent noise in some ways. For one, ADAPT-VQE runs with coherent noise and error mitigation had more inflated gradient norms than runs with incoherent noise and error mitigation. Figures~\ref{fig:hm-01-allcoh-pzne} and \ref{fig:hm-03-allcoh-pzne} show that the combination of PT and ZNE against coherent noise inflated gradient magnitudes even more compared to other methods.  Additionally, error mitigation used against coherent noise affected the gradient magnitudes of all operators in the pool, contrasting with the case of incoherent noise, where only a subset of operators was affected. Just as in the case of incoherent noise, error mitigation did not completely recreate the noiseless ansatz, although it was successful in achieving the ultimate goal of finding a chemically-accurate solution.

\section{Conclusion} \label{Section V}

In this work, we studied the impact of noise on the operator selection process of ADAPT-VQE. Both incoherent and coherent noise channels were simulated with varying strengths. A useful characteristic of ADAPT-VQE is that effective operators for the ansatz don't need to have the highest gradient magnitude; this study confirmed this by showing that effective operators are still appended to the ansatz in the presence of low levels of noise. However, for higher noise strengths, both incoherent and coherent noise channels often pushed ADAPT-VQE to select ineffective operators, preventing the algorithm from reaching a chemically accurate solution to the ground state energy of H$_3$. Additionally, both types of channels caused the gradient landscape to stall. This led the algorithm to repeatedly select the same operator, effectively preventing convergence. 

To assess whether EM techniques can compensate for noise and allow ADAPT-VQE to converge to the solution, we implemented a suite of well-known methods---dynamical decoupling, zero-noise extrapolation, and Pauli-twirling---and analyzed their impact on noisy ADAPT-VQE runs that would otherwise not converge. We observed that these techniques are able to improve the operator selection and delay or prevent the stalling of the gradients, emabling the algorithm to reach a chemically-accurate solution.

In the case of incoherent noise channels, the combination of dynamical decoupling and zero-noise extrapolation stood out as the most effective EM strategy. Using both simultaneously prevented the gradient landscape from flattening, allowed a different operator to be selected at each iteration, and restored ADAPT-VQE’s ability to converge to a chemically accurate ground state energy for the H$_3$ molecule. By contrast, applying these methods individually only worked for a subset of the incoherent noise channels against which the combination was effective. Coherent noise required a different recipe. We observed that Pauli twirling was effective when combined with dynamical decoupling, zero-noise extrapolation, or both. 

We note that, in general, error mitigation did not reproduce the noiseless ansatz, as gradient landscapes were very different from the noiseless case, even for runs that successfully converged to the desired accuracy. This is not a flaw of the techniques, rather a testament to the flexibility of the ADAPT-VQE selection criterion: even in the presence of noise, the algorithm is able to reach chemical accuracy in spite of the ansatz diverging from the one that would be generated in the absence of noise. 

However, in general, we observed that noise causes the inflation of gradient magnitudes to a degree that prevents the algorithm from terminating under the standard gradient-magnitude-based convergence criterion. Furthermore, EM techniques often aggravate this issue, particularly so in the presence of coherent noise, and sometimes beyond the inflation caused by the noise itself. This suggests that the convergence criterion may require adjustments to be viable in a realistic setting.

Our work provides insights as to how noise impacts the ADAPT-VQE selection criterion and to what degree this can be improved by EM techniques. Interesting directions for future research include developing a noise-resilient selection criterion, generalizing our study to larger system sizes, and implementing the algorithm on quantum hardware.

\section{Acknowledgments}

We thank Sophia Economou for helpful discussions on this work. This work was supported by NSF grant No. 2231328.

\bibliography{ADAPT-VQE.bib/ADAPT-VQE.bib.bib}

\clearpage
\appendix

\section{Incoherent Noise}\label{Section VI A}
This appendix presents gradient heatmaps and energy error plots from ADAPT-VQE runs subject to amplitude damping, phase damping, and depolarizing noise. For the sake of succinctness, only plots at noise strength 0.1 are shown; unless stated otherwise, it can be assumed that all trends listed in this appendix are also applicable to the runs with noise strength 0.3.

\begin{figure*}[htbp]
  \centering

  \begin{subfigure}[t]{0.49\textwidth}
    \centering
    \includegraphics[width=\linewidth]{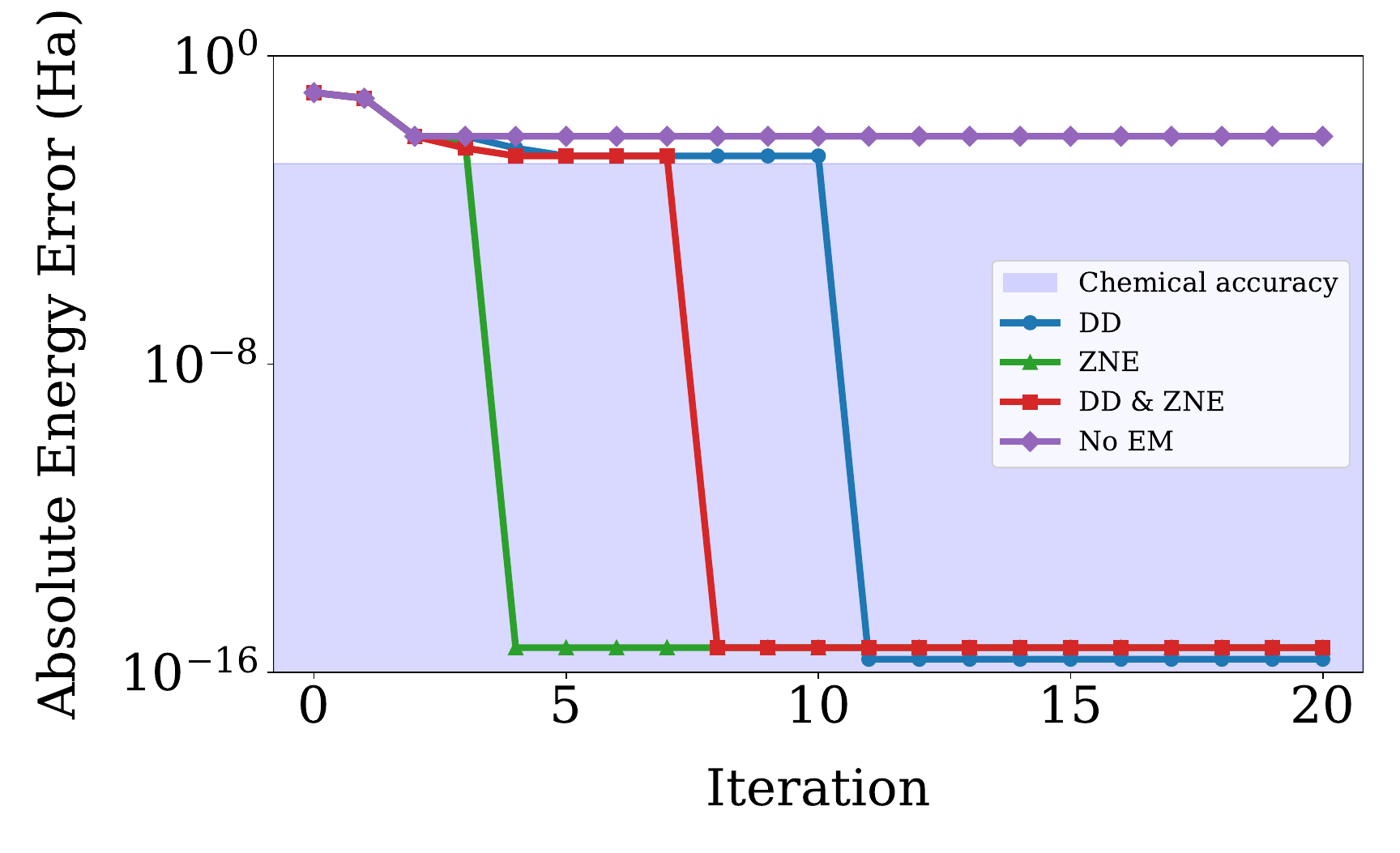}
    \caption{\centering Amplitude Damping (0.1)}
    \label{fig:ee-ad}
  \end{subfigure}

  \vspace{0.6em}

  \begin{subfigure}[t]{0.49\textwidth}
    \centering
    \includegraphics[width=\linewidth]{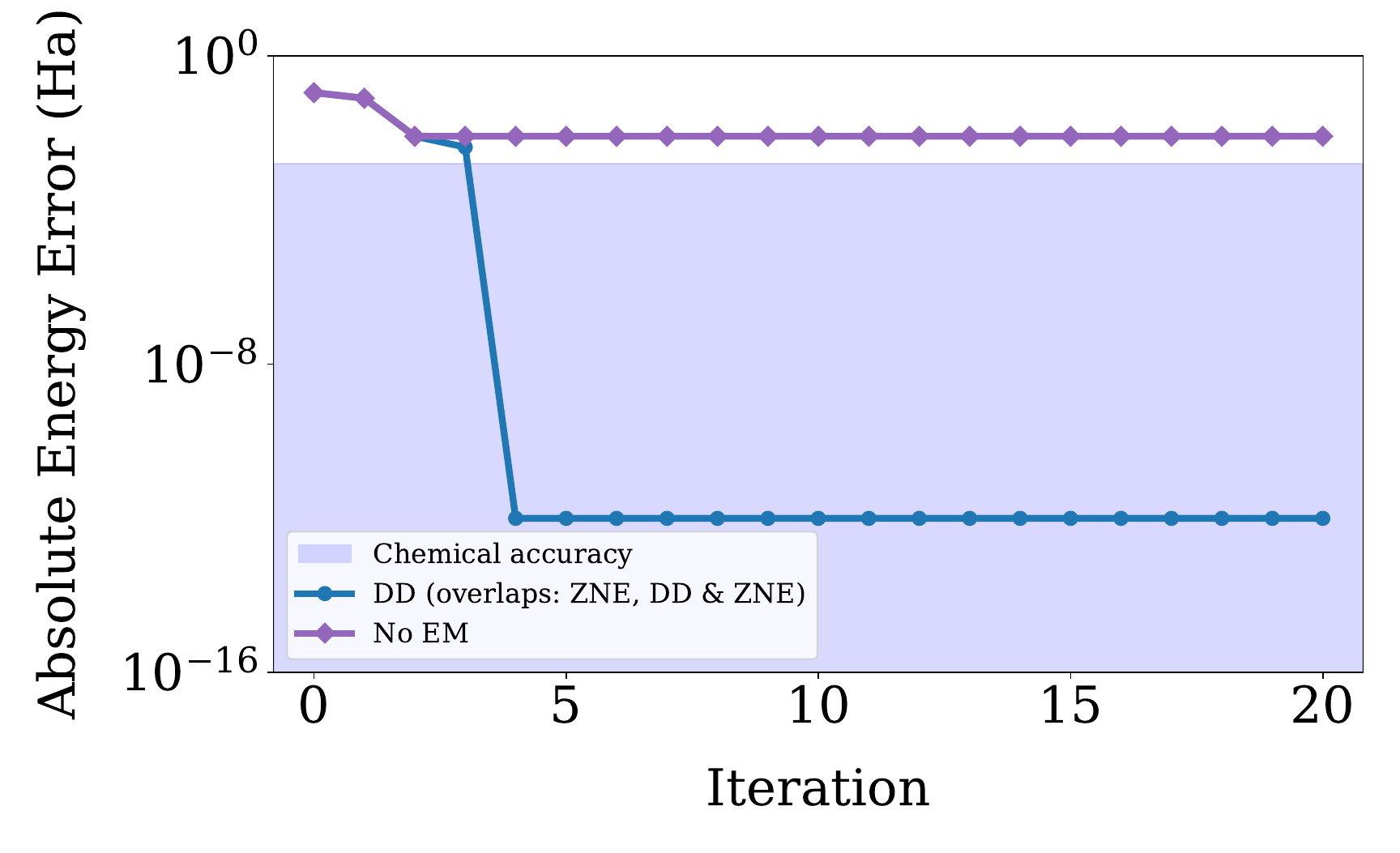}
    \caption{\centering Phase Damping (0.1)}
    \label{fig:ee-pd}
  \end{subfigure}\hfill
  \begin{subfigure}[t]{0.49\textwidth}
    \centering
    \includegraphics[width=\linewidth]{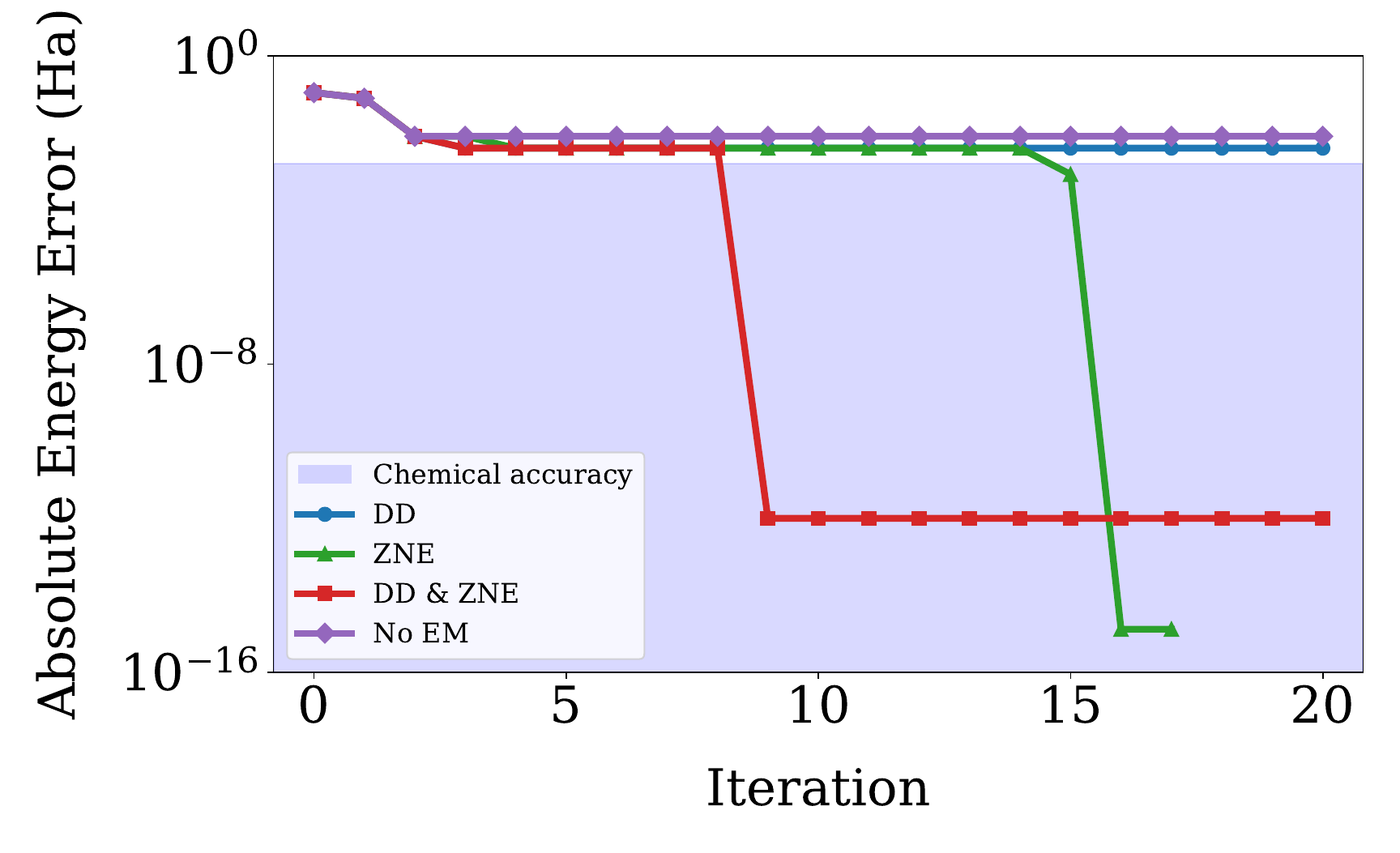}
    \caption{\centering Depolarizing Noise (0.1)}
    \label{fig:ee-dn}
  \end{subfigure}

  \caption{Energy error plots for all mitigation methods across the three incoherent noise channels at strength 0.1.}
  \label{fig:appendix-group-energy-error}
\end{figure*}

As with the `all incoherent' channel discussed in the main text, all three incoherent noise channels prevented ADAPT-VQE from reaching a chemically-accurate ground energy. This is illustrated by the energy error plots in Fig.~\ref{fig:appendix-group-energy-error}. However, for all 3 noise channels, some form of EM was able to restore ADAPT-VQE's ability to reach chemical accuracy. As with the `all incoherent' channel, the combination of DD and ZNE proved to be the most effective error-mitigation strategy, as it enabled ADAPT-VQE to reach chemical accuracy across all tested incoherent noise channels. DD and ZNE used separately were only successful in a subset of cases. In Fig.~\ref{fig:ee-dn}, for example, DD was unsuccessful at mitigating the effects of depolarizing noise. Though not illustrated in this appendix, ZNE used individually was unsuccessful at mitigating the effects of various noise channels at the 0.3 noise level.

\begin{figure*}[htbp]
  \centering

  \begin{subfigure}[t]{0.48\textwidth}
    \centering
    \includegraphics[width=\linewidth]{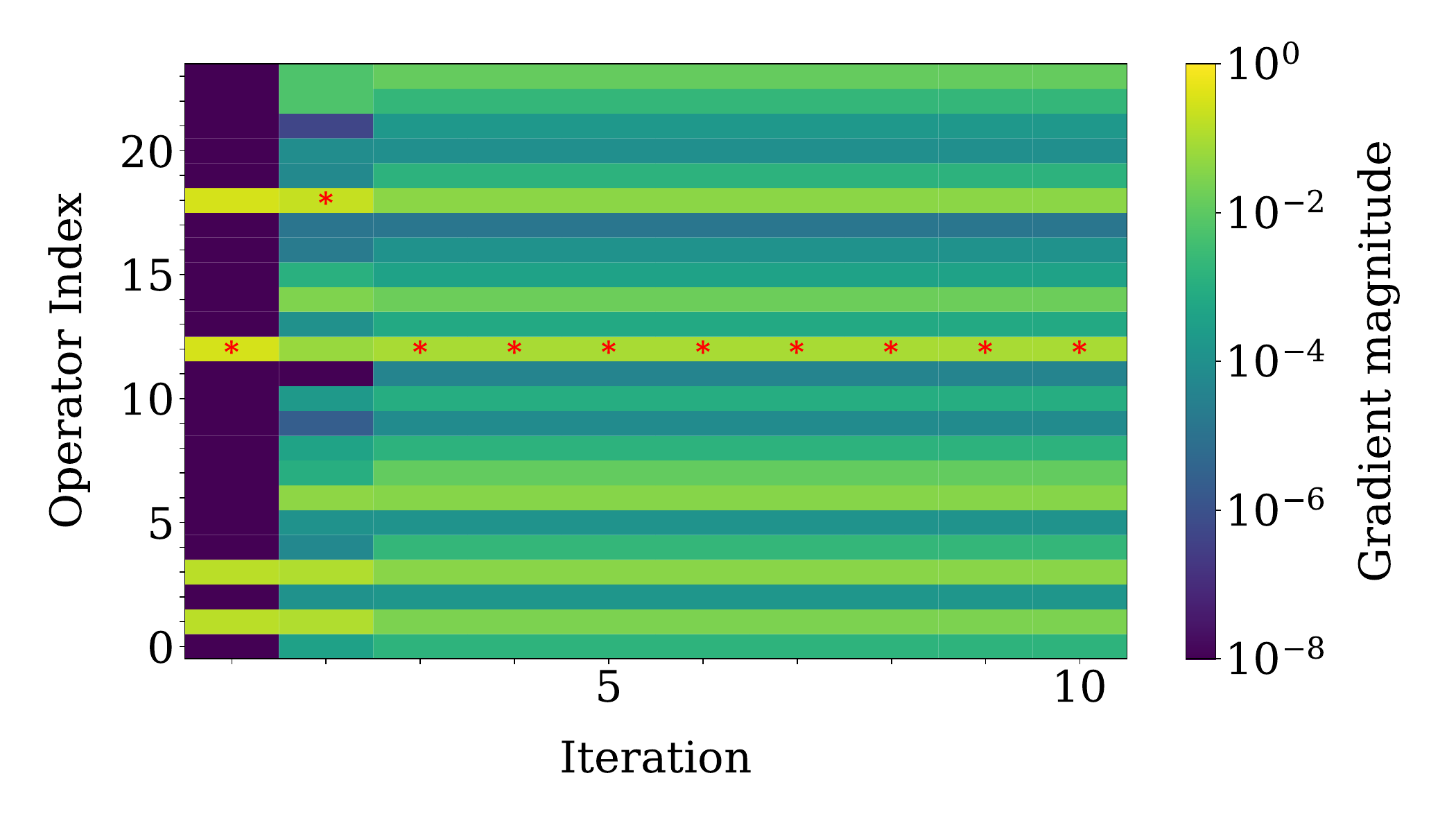}
    \caption{\centering No EM: Amplitude Damping (0.1)}
    \label{fig:hm-ad-noem}
  \end{subfigure}\hfill
  \begin{subfigure}[t]{0.48\textwidth}
    \centering
    \includegraphics[width=\linewidth]{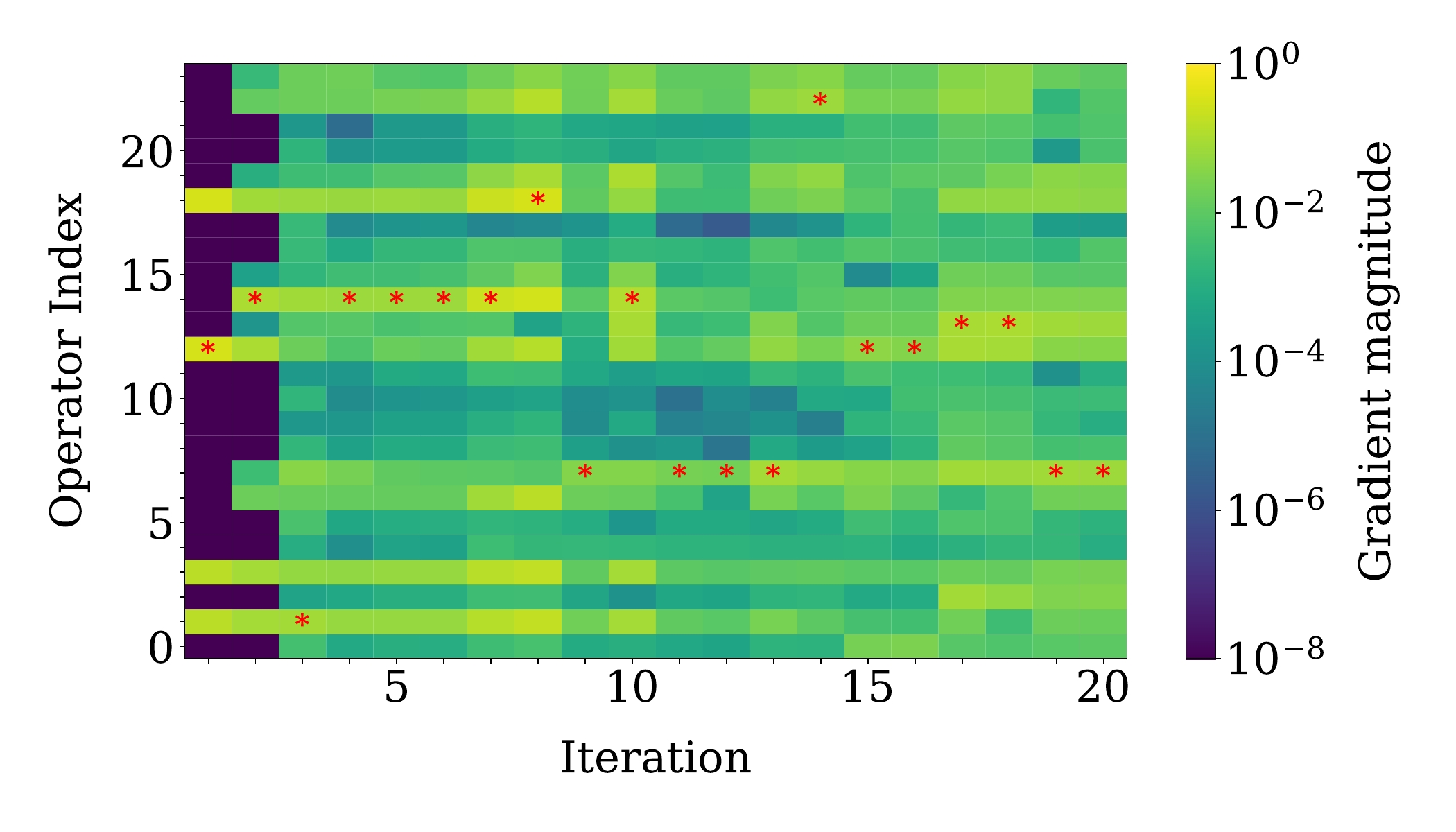}
    \caption{\centering DD \& ZNE: Amplitude Damping (0.1)}
    \label{fig:hm-ad-ddzne}
  \end{subfigure}

  \vspace{1em}

  \begin{subfigure}[t]{0.48\textwidth}
    \centering
    \includegraphics[width=\linewidth]{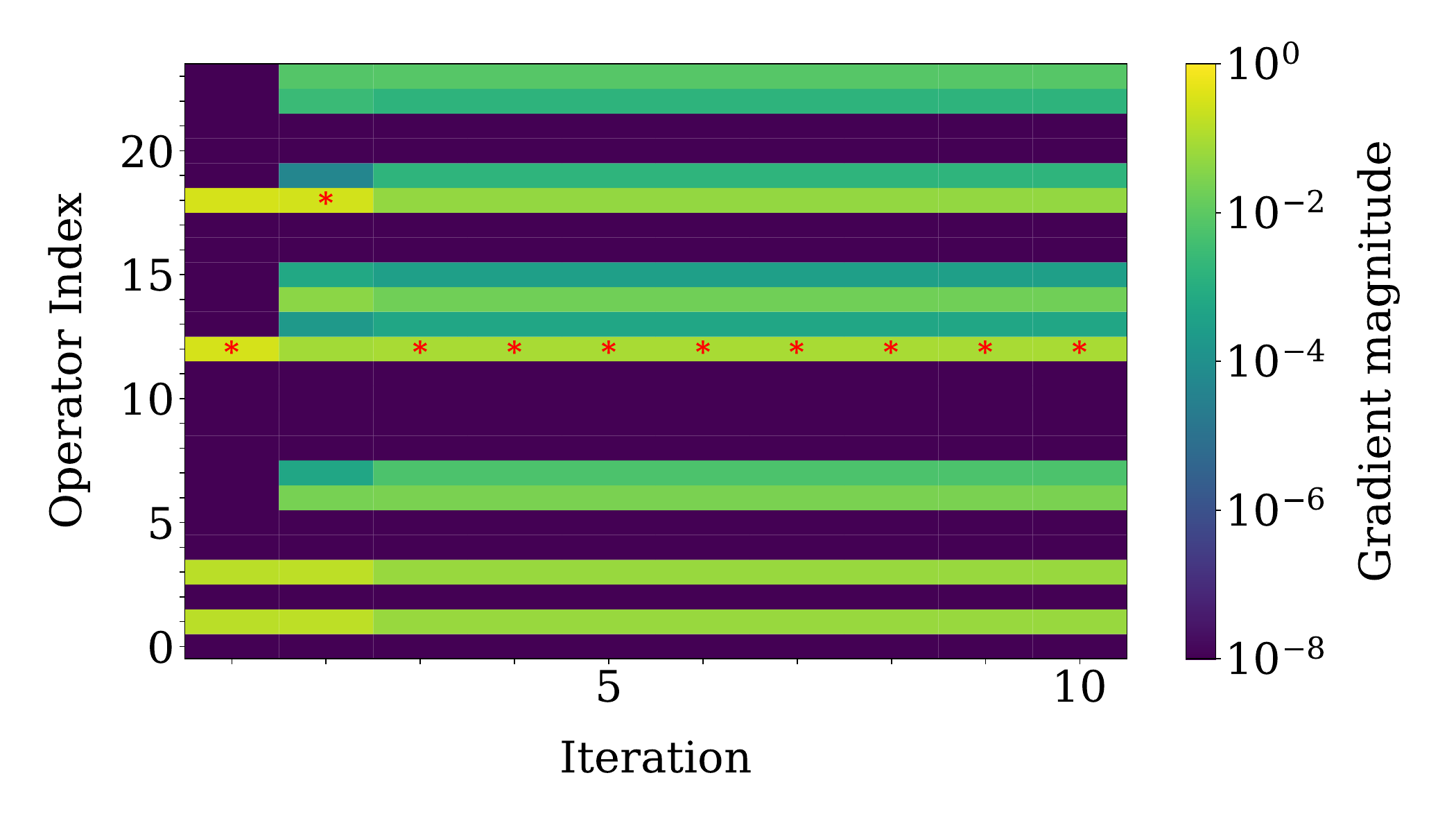}
    \caption{\centering No EM: Phase Damping (0.1)}
    \label{fig:hm-pd-noem}
  \end{subfigure}\hfill
  \begin{subfigure}[t]{0.48\textwidth}
    \centering
    \includegraphics[width=\linewidth]{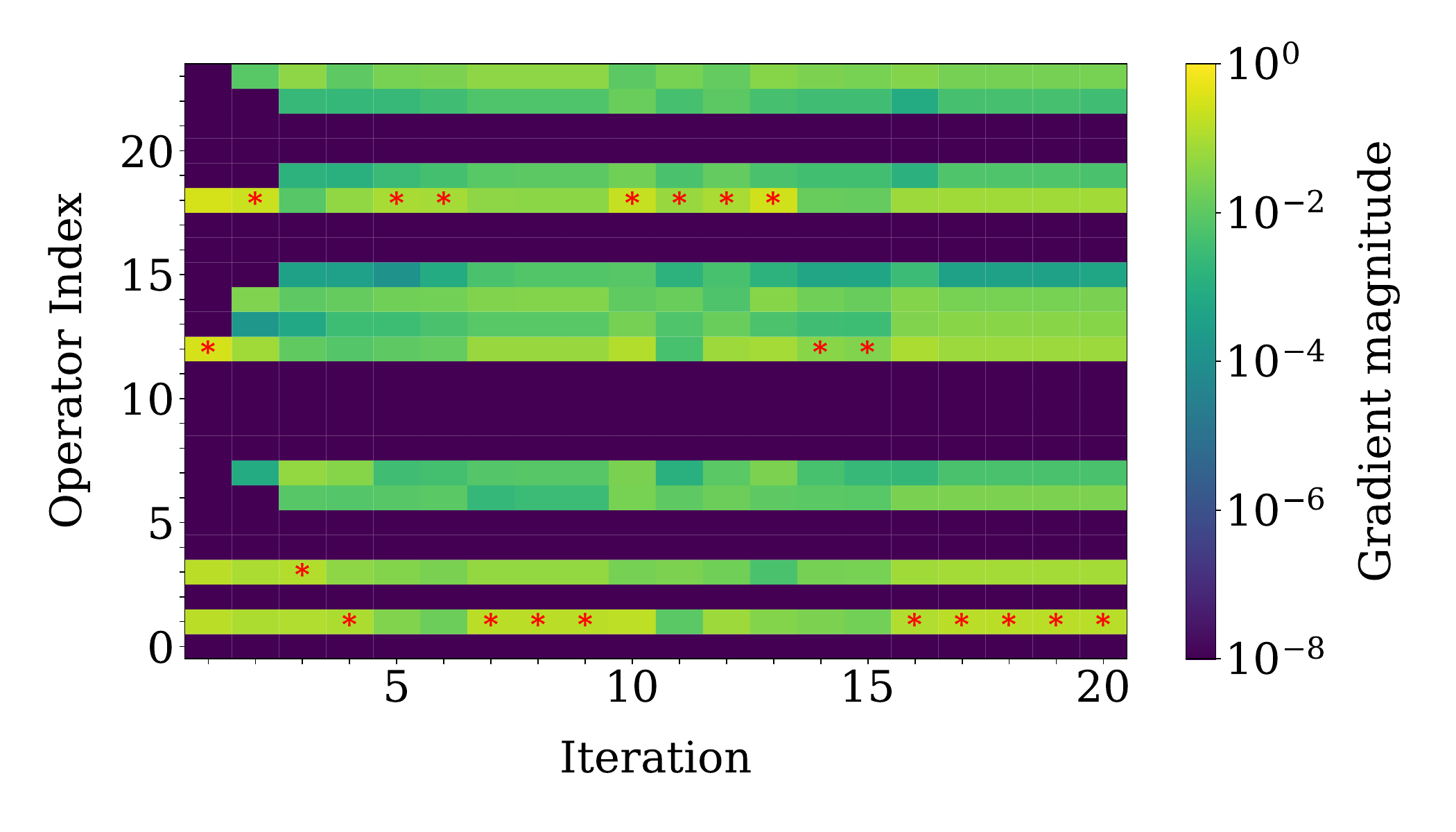}
    \caption{\centering DD \& ZNE: Phase Damping (0.1)}
    \label{fig:hm-pd-ddzne}
  \end{subfigure}

  \vspace{1em}

  \begin{subfigure}[t]{0.48\textwidth}
    \centering
    \includegraphics[width=\linewidth]{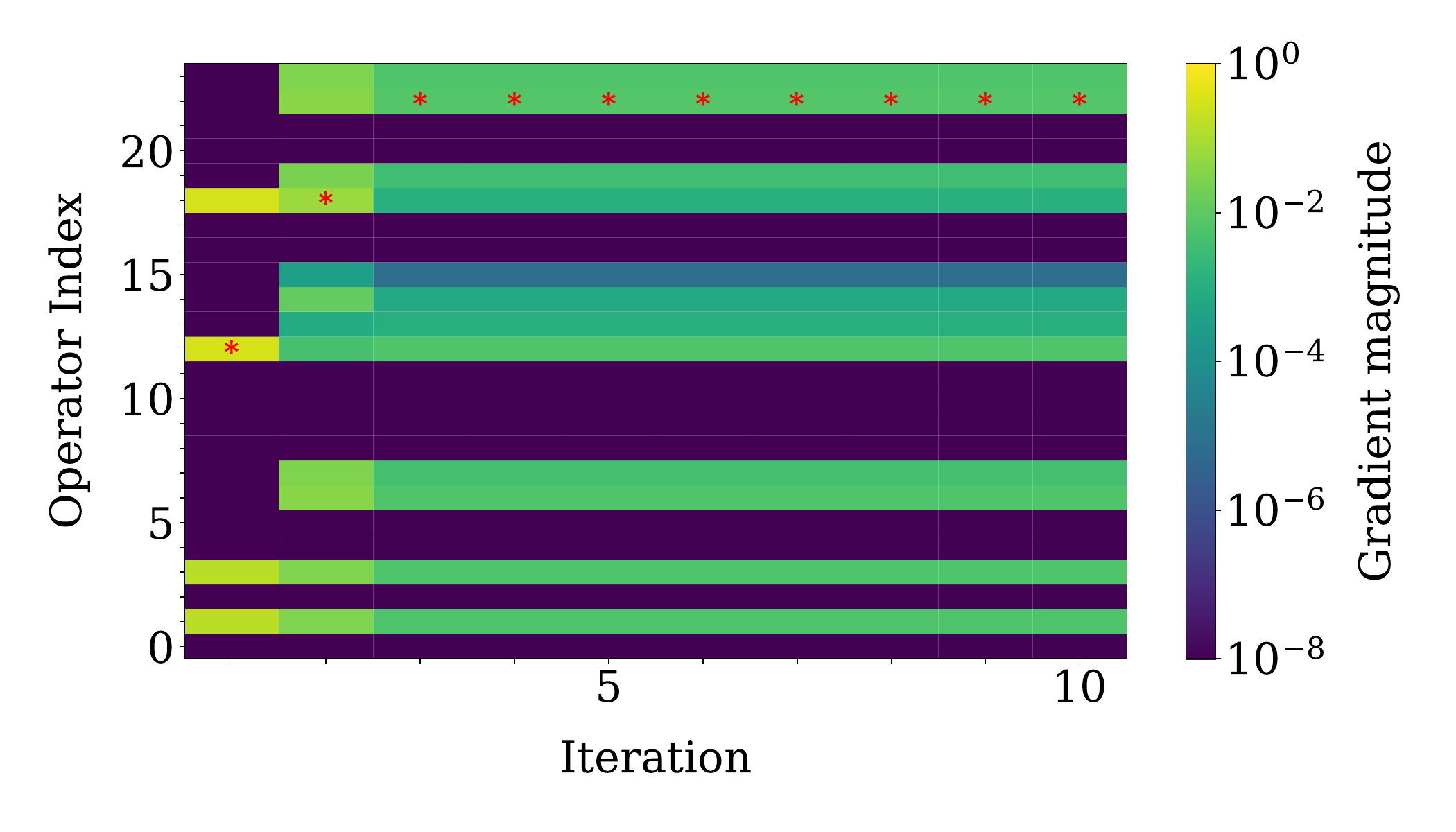}
    \caption{\centering No EM: Depolarizing Noise (0.1)}
    \label{fig:hm-dn-noem}
  \end{subfigure}\hfill
  \begin{subfigure}[t]{0.48\textwidth}
    \centering
    \includegraphics[width=\linewidth]{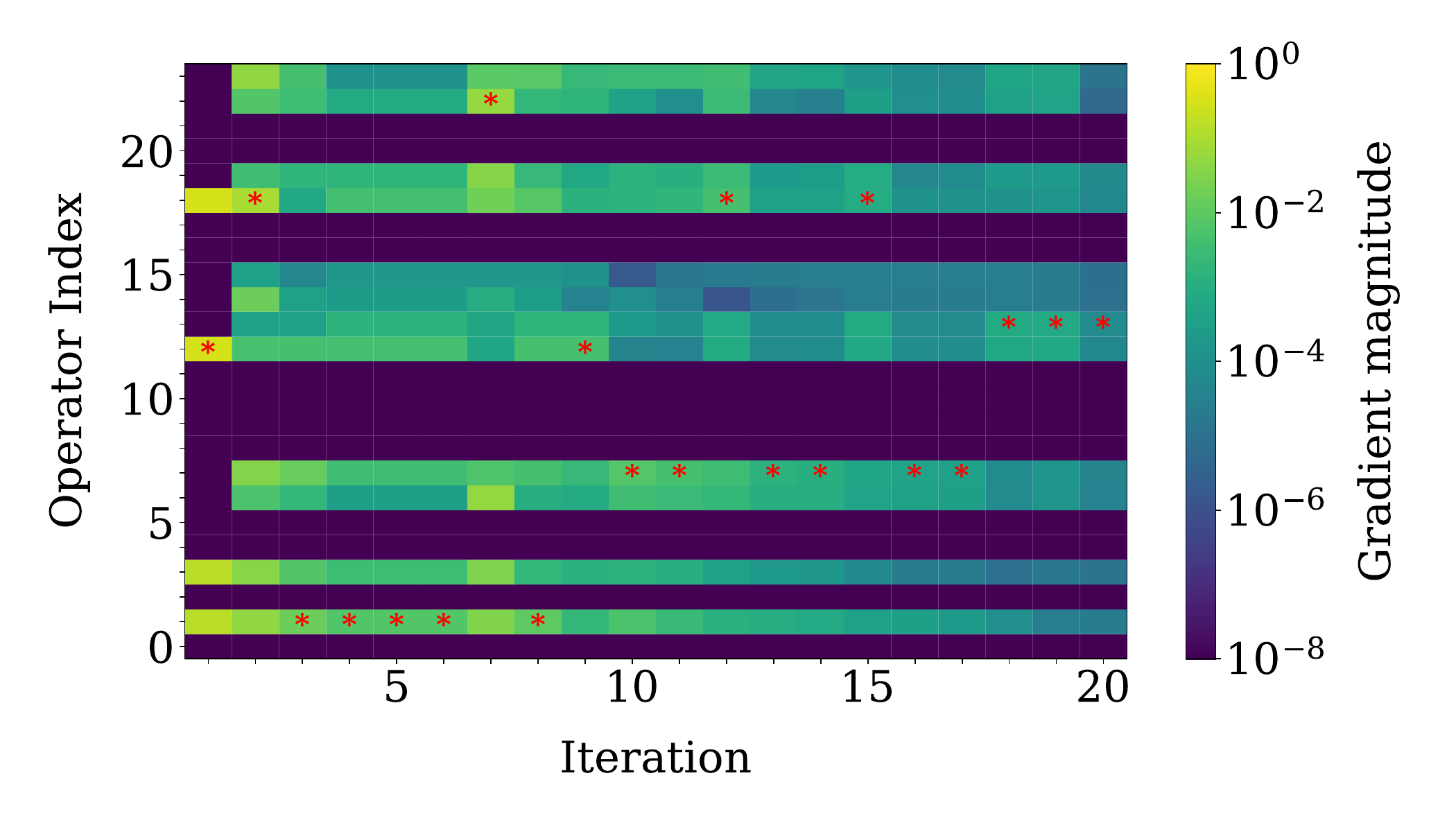}
    \caption{\centering DD \& ZNE: Depolarizing Noise (0.1)}
    \label{fig:hm-dn-ddzne}
  \end{subfigure}

  \caption{Gradient heatmaps without error mitigation (left column) and with DD \& ZNE (right column) across the three incoherent noise channels at strength 0.1. A red star on an operator block indicates that it was selected to be appended to the ansatz.}
  \label{fig:appendix-combined-heatmaps}
\end{figure*}

 The effects of amplitude damping, phase damping, and depolarizing noise on the gradient landscape were also very similar to the `all incoherent' channel: as shown by the left column of Fig.~\ref{fig:appendix-combined-heatmaps}, these noise channels stagnated the gradient landscape and led to repeated operator selection. Among these, amplitude damping  (Fig.~\ref{fig:hm-ad-noem}) was unique in that it caused a larger percentage of the operator pool to have non-zero gradients throughout the run, similar to coherent noise. When EM was applied to these noisy runs, the resulting behavior again closely resembled that observed for the `all incoherent' channel: when EM was successful, it prevented the gradient landscape from stagnating over the 20 iterations tested and allowed operator selection to vary across iterations. As shown in the right column of Fig.~\ref{fig:appendix-combined-heatmaps}, this behavior is consistent across all three incoherent noise channels. It should be noted that, like the `all incoherent' channel, EM did not completely restore the noiseless ansatz. Again, this was not a limitation to EM as several operators produce significant energy changes, so the exact ansatz does not need to be conserved. Although not shown in this appendix, when EM was unsuccessful at restoring chemical accuracy, the stagnation of the gradient landscape was only postponed rather than prevented during the run. This behavior represents another characteristic shared with the `all incoherent' noise channel and is described in more detail in the main text.
 
\section{Coherent Noise} \label{Section VI B}

This appendix presents gradient heatmaps and energy error plots from ADAPT-VQE runs subject to $X$ and $Z$ over-rotations. Again, for the sake of succinctness, only plots at the 0.1 noise level are shown; unless stated otherwise, it can be assumed that all trends listed in this appendix are applicable to the 0.3 noise level. 

\begin{figure*}[htbp]
\centering
  \begin{subfigure}[t]{0.48\textwidth}
    \centering
    \includegraphics[width=\linewidth]{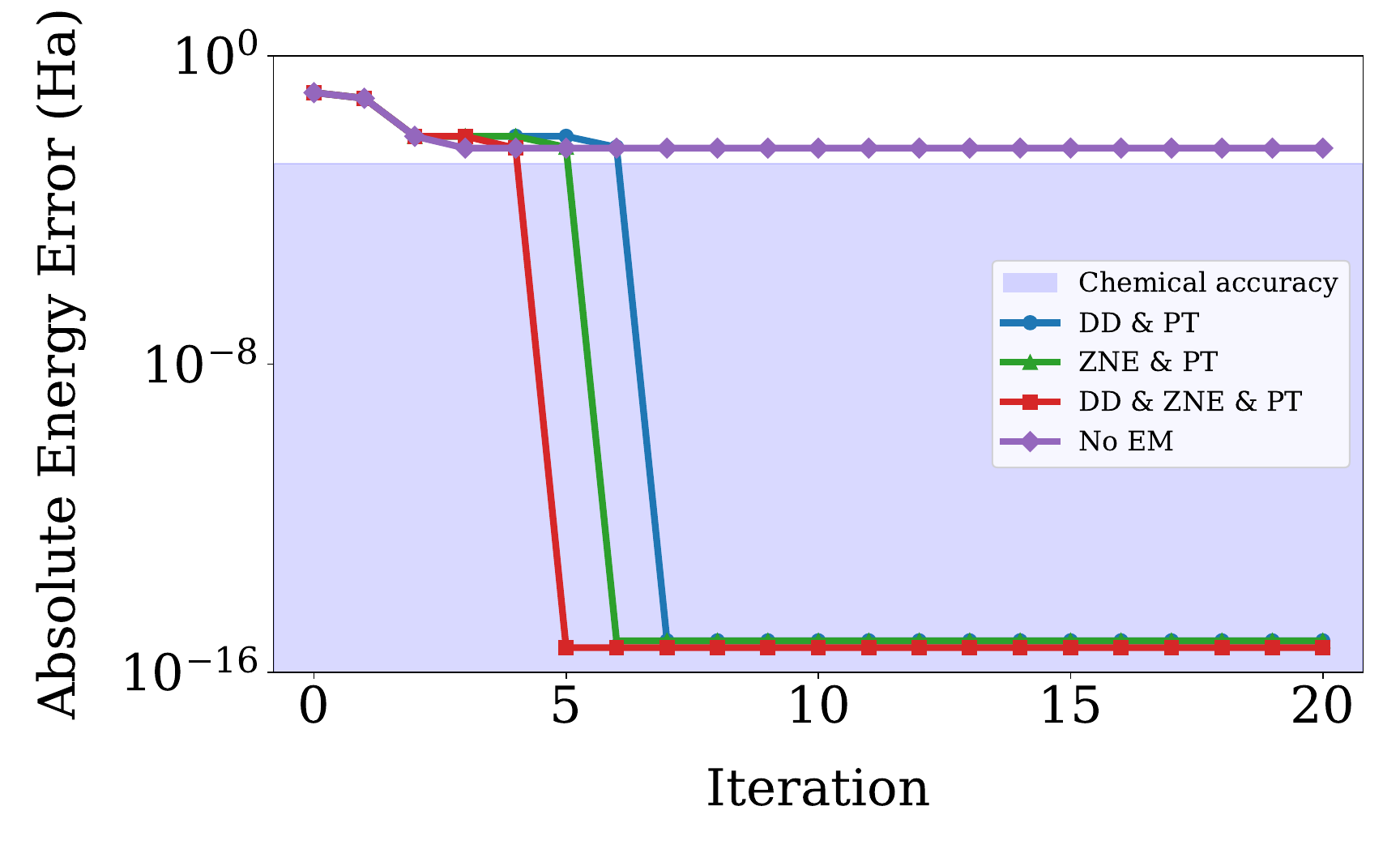}
    \caption{\centering $X$ over-rotation (0.1)}
    \label{fig:ee-x-01}
  \end{subfigure}
  \hfill
  \begin{subfigure}[t]{0.48\textwidth}
    \centering
    \includegraphics[width=\linewidth]{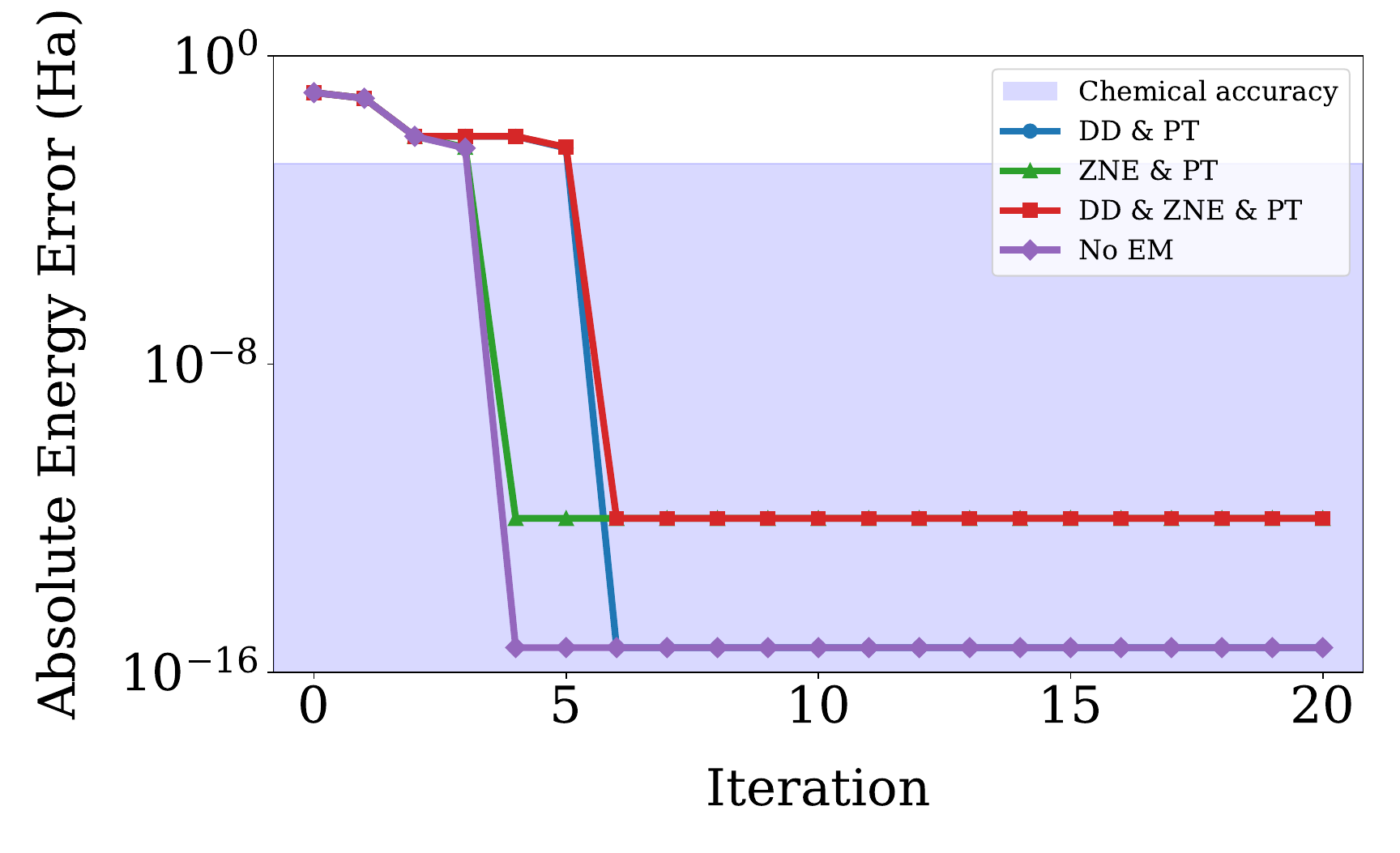}
    \caption{\centering $Z$ over-rotation (0.1)}
    \label{fig:ee-z-01}
  \end{subfigure}
  \caption{Energy error plots for all mitigation methods across the three coherent noise channels at strength 0.1.}
  \label{fig:appendix_ee_coherent_01}
\end{figure*}

Unlike the `all coherent' channel, $X$ and $Z$ over-rotation had varying effects on ADAPT-VQE's ability to reach a chemically-accurate ground energy. As shown in Fig.~\ref{fig:appendix_ee_coherent_01}, the $X$ over-rotations prevented ADAPT-VQE from reaching chemical accuracy, while the $Z$ over-rotations had no effect on chemical accuracy. At the 0.3 noise level, both coherent noise channels prevented ADAPT-VQE from reaching chemical accuracy. While not explored in detail, the ability of ADAPT-VQE to reach chemical accuracy under the 0.1 $Z$ over-rotation channel suggests a degree of inherent robustness to this noise source.

\begin{figure*}[htbp]
\centering

  \begin{subfigure}[t]{0.48\textwidth}
    \centering
    \includegraphics[width=\linewidth]{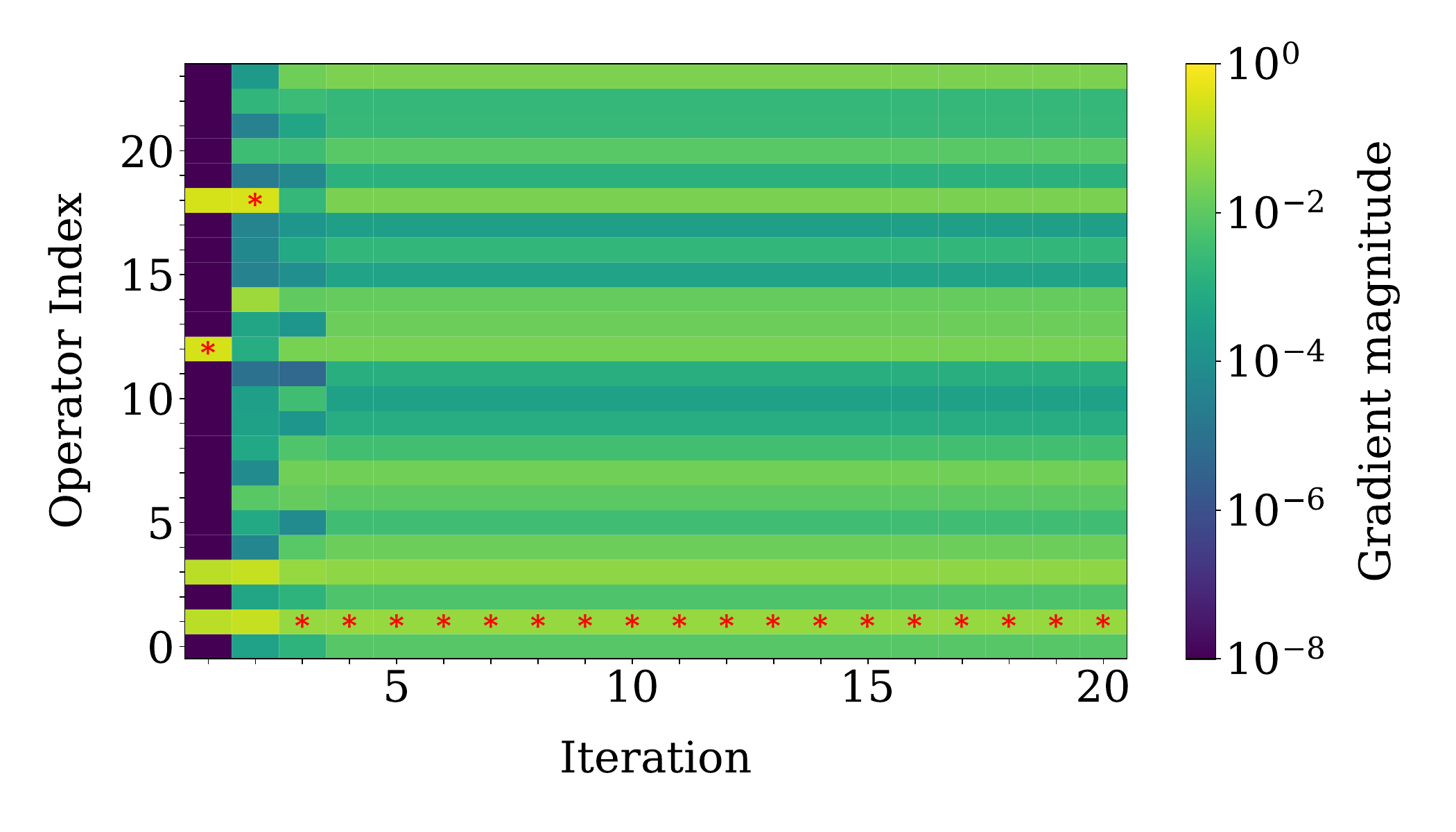}
    \caption{\centering No EM: $X$ over-rotation (0.1)}
    \label{fig:hm-x-noEM-01}
  \end{subfigure}
  \hfill
  \begin{subfigure}[t]{0.48\textwidth}
    \centering
    \includegraphics[width=\linewidth]{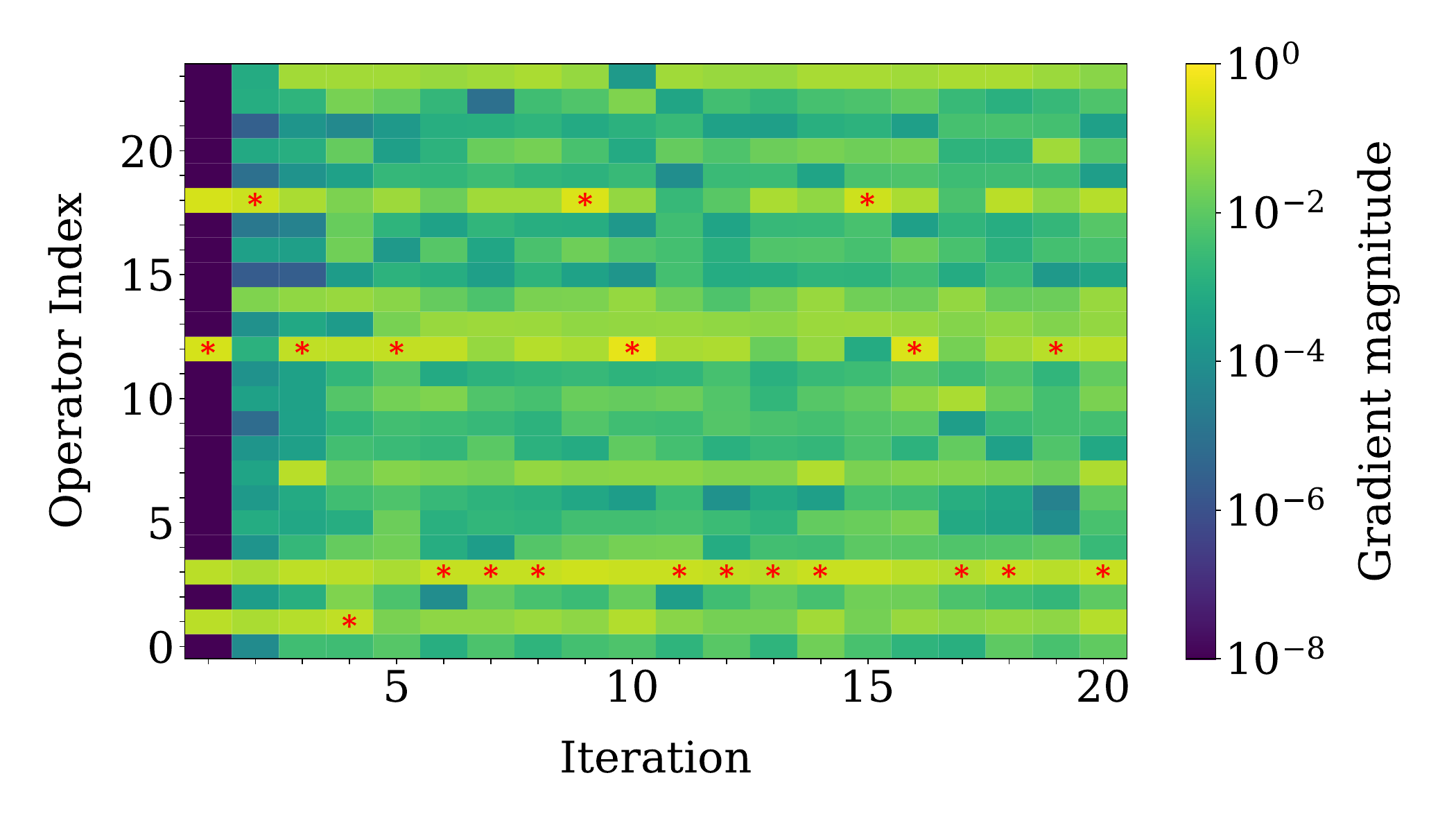}
    \caption{\centering DD \& ZNE \& PT: $X$ over-rotation (0.1)}
    \label{fig:hm-x-ddpzne-01}
  \end{subfigure}

  \vspace{1.0em}

  \begin{subfigure}[t]{0.48\textwidth}
    \centering
    \includegraphics[width=\linewidth]{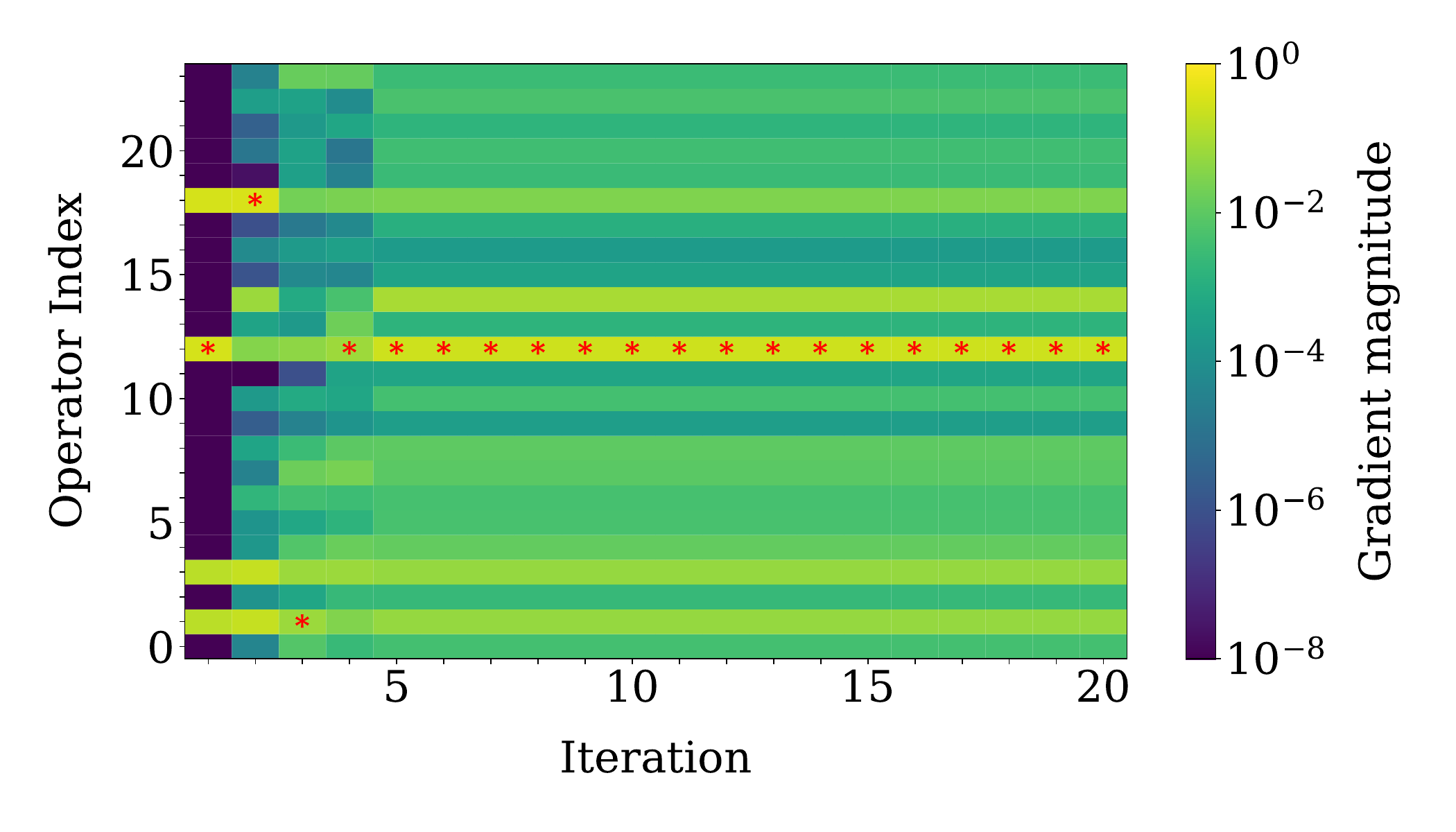}
    \caption{\centering No EM: $Z$ over-rotation (0.1)}
    \label{fig:hm-z-noEM-01}
  \end{subfigure}
  \hfill
  \begin{subfigure}[t]{0.48\textwidth}
    \centering
    \includegraphics[width=\linewidth]{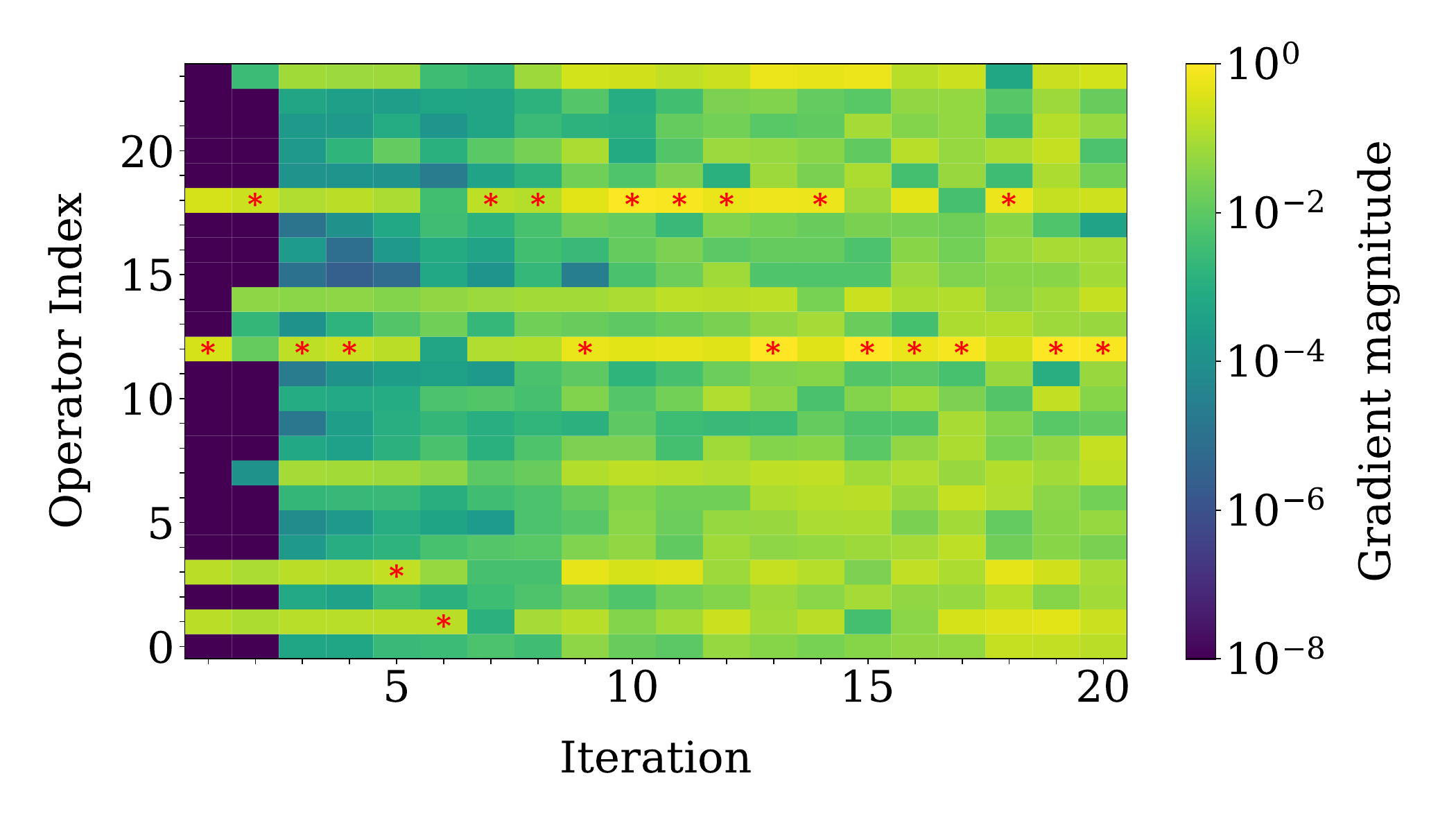}
    \caption{\centering DD \& ZNE \& PT: $Z$ over-rotation (0.1)}
    \label{fig:hm-z-ddpzne-01}
  \end{subfigure}

  \caption{Gradient heatmaps without error mitigation (left column) and with DD \& ZNE \& PT (right column) across the three coherent noise channels discussed in the main text at a noise level of 0.1.  A red star on an operator block indicates that it was selected to be appended to the ansatz.}
  \label{fig:appendix_hm_comparison_01}
\end{figure*}

Under all tested noise conditions, the applied EM methods enabled ADAPT-VQE to recover or maintain chemical accuracy. Additionally, unlike the behavior observed under incoherent noise channels, all three EM configurations proved consistently effective across all simulated runs. This is most effectively illustrated by Fig.~\ref{fig:ee-x-01}, where DD \& PT, DD \& ZNE, and DD \& PT \& ZNE all allowed ADAPT-VQE to reach chemical accuracy. These results suggest that PT, when paired with DD, ZNE, or both, serves as an effective strategy for countering the coherent noise channels investigated in this work.

The effects of $X$ and $Z$ over-rotation on the gradient landscape were very similar to the `all coherent' channel discussed in the main text: As shown by the left column of Fig.~\ref{fig:appendix_hm_comparison_01}, these noise channels stagnated operator gradients, led to the selection of repeated operators, and greatly increased the number of operators with non-zero gradients. The impact of EM under the $X$ and $Z$ over-rotation channels was also similar to the impact under the `all coherent' channels: In all tested cases, EM prevented the stagnation of the gradient landscape, allowed for different operators to be selected at each iteration, and improved gradient magnitudes. These results are illustrated by the right column of Fig.~\ref{fig:appendix_hm_comparison_01}.

\end{document}